\documentclass[prc,twocolumn,twoside,showpacs,nofootinbib,floatfix]{revtex4}

\usepackage{graphicx,color}
\usepackage{amsmath,amssymb,bm}
\usepackage{ae}

\newcommand{\eq}[1]{\begin{equation}#1\end{equation}}
\newcommand{\eqmulti}[1]{\begin{equation}\begin{split}#1\end{split}\end{equation}}

\newcommand{\ket}[1]{\ensuremath{\,|{#1}\rangle}}

\newcommand{\braket}[2]{\ensuremath{\langle{#1}|{#2}\rangle}}

\newcommand{\matrixe}[3]{\ensuremath{\langle{#1}|\,{#2}\,|{#3}\rangle}}

\newcommand{\expect}[1]{\ensuremath{\langle{#1}\rangle}}

\newcommand{\op}[1]{\ensuremath{\mathrm{#1}}}
\newcommand{\adj}[1]{\ensuremath{{{#1}}^{\dag}}}
\newcommand{\corr}[1]{\ensuremath{\widetilde{#1}}}

\newcommand{\ii}{\ensuremath{\mathrm{i}}}
\newcommand{\dd}{\ensuremath{\mathrm{d}}}

\renewcommand{\vec}[1]{\ensuremath{\bm{#1}}}

\newcommand{\sixjsymb}[6]{\ensuremath{\left\{ \begin{matrix} 
  #1 & #2 & #3 \\ #4 & #5 & #6 \end{matrix} \right\}}}
\newcommand{\ninejsymb}[9]{\ensuremath{\left\{\begin{matrix} 
  #1 & #2 & #3 \\ #4 & #5 & #6 \\ #7 & #8 & #9 \end{matrix}\right\}}}
\newcommand{\talmimosh}[9]{\ensuremath{\langle\!\langle#1 #2, #3 #4 \,|\, #5 #6,
#7 #8 ; #9\rangle\!\rangle}}

\newcommand{\cO}{\ensuremath{\op{c}}}
\newcommand{\ccO}{\ensuremath{\adj{\op{c}}}}
\newcommand{\gO}{\ensuremath{\op{g}}}

\newcommand{\qO}{\ensuremath{\op{q}}}
\newcommand{\rO}{\ensuremath{\op{r}}}

\newcommand{\vO}{\ensuremath{\op{v}}}

\newcommand{\AO}{\ensuremath{\op{A}}}
\newcommand{\CO}{\ensuremath{\op{C}}}
\newcommand{\CCO}{\ensuremath{\adj{\op{C}}}}

\newcommand{\HO}{\ensuremath{\op{H}}}

\newcommand{\RO}{\ensuremath{\op{R}}}

\newcommand{\TO}{\ensuremath{\op{T}}}
\newcommand{\VO}{\ensuremath{\op{V}}}

\newcommand{\PiO}{\ensuremath{\op{\Pi}}}

\newcommand{\rV}{\ensuremath{\vec{r}}}

\newcommand{\AC}{\ensuremath{\mathcal{A}}}

\newcommand{\pOV}{\ensuremath{\vec{\op{p}}}}
\newcommand{\qOV}{\ensuremath{\vec{\op{q}}}}
\newcommand{\rOV}{\ensuremath{\vec{\op{r}}}}

\newcommand{\xOV}{\ensuremath{\vec{\op{x}}}}

\newcommand{\LOV}{\ensuremath{\vec{\op{L}}}}

\newcommand{\XOV}{\ensuremath{\vec{\op{X}}}}

\newcommand{\sigmaOV}{\ensuremath{\vec{\op{\sigma}}}}

\newcommand{\tensorRQO}{\ensuremath{\op{s}_{12}(\rOV,\qOV_{\Omega})}}

\newcommand{\Rm}{\ensuremath{R_-}}
\newcommand{\DRm}{\ensuremath{R'_-}}

\newcommand{\Rp}{\ensuremath{R_+}}

\newcommand{\Rpm}{\ensuremath{R_{\pm}}}
\newcommand{\Rmp}{\ensuremath{R_{\mp}}}

\newcommand{\half}{\ensuremath{\tfrac{1}{2}}}

\newcommand{\UCOM}{\ensuremath{\textrm{UCOM}}}
\newcommand{\intr}{\ensuremath{\textrm{int}}}

\newcommand{\cm}{\ensuremath{\textrm{cm}}}
\newcommand{\elem}[2]{\ensuremath{{}^{#2}\text{#1}}}

\begin{document}

\title{Hartree-Fock and Many-Body Perturbation Theory with\\ Correlated Realistic NN-Interactions}

\author{R. Roth}
\email{robert.roth@physik.tu-darmstadt.de}

\author{P. Papakonstantinou} 
\author{N. Paar}
\author{H. Hergert}

\affiliation{Institut f\"ur Kernphysik, Technische Universit\"at Darmstadt,
64289 Darmstadt, Germany}

\author{T. Neff}

\affiliation{National Superconducting Cyclotron Laboratory, Michigan State University, 
East Lansing, Michigan 48824, USA}

\author{H. Feldmeier}

\affiliation{Gesellschaft f\"ur Schwerionenforschung, Planckstr. 1, 
64291 Darmstadt, Germany}

\date{\today}

\begin{abstract}    
We employ correlated realistic nucleon-nucleon interactions for the description of nuclear ground states throughout the nuclear chart within the Hartree-Fock approximation. The crucial short-range central and tensor correlations, which are induced by the realistic interaction and cannot be described by the Hartree-Fock many-body state itself, are included explicitly by a state-independent unitary transformation in the framework of the unitary correlation operator method (UCOM). Using the correlated realistic interaction $\VO_{\UCOM}$ resulting from the Argonne V18 potential, bound nuclei are obtained already on the Hartree-Fock level. However, the binding energies are smaller than the experimental values because long-range correlations have not been accounted for. Their inclusion by means of many-body perturbation theory leads to a remarkable agreement with experimental binding energies over the whole mass range from \elem{He}{4} to \elem{Pb}{208}, even far off the valley of stability. The observed perturbative character of the residual long-range correlations and the apparently small net effect of three-body forces provides promising perspectives for a unified nuclear structure description.
\end{abstract}


\maketitle

\clearpage
\section{Introduction}
\label{sec:intro}

The description of heavier nuclei starting from realistic nucleon-nucleon (NN) interactions which reproduce the experimental NN phase shifts is a long-standing and unsolved problem. So far, the theoretical tools applicable in the mass region beyond $A\approx60$ are predominantly density-functional approaches based on purely phenomenological energy functionals. Recent developments aim at more fundamental energy functionals motivated from and constrained by QCD \cite{FiKa04}. At the same time, light nuclei have been treated very successfully in so-called ab-initio approaches, e.g. Green's function Monte Carlo \cite{PiWi01, PiWi04, PiVa02} or no-core shell model \cite{CaNa02,NaOr02,NaVa00}. These calculations have shown that realistic NN-interactions, supplemented by a three-body force, are able to describe the nuclear structure of light isotopes quite well. However, due to their computational complexity, these practically exact numerical solutions of the quantum many-body problem cannot be applied to nuclei beyond the $p$-shell. 

Strong correlations in the many-body system are the basic issue which has to be addressed when starting from a realistic NN interaction. The naive use of a bare realistic potential, e.g. Argonne V18 (AV18) \cite{WiSt95} or CD Bonn \cite{Mach01}, in a simple many-body approximation like Hartree-Fock (HF) will not lead to sensible results. The many-body states of the HF approximation are Slater determinants, i.e., independent particle states incapable of describing any correlations. 

Already the deuteron elucidates the nature of the dominant interaction-induced correlations. The relative two-body wave function shows a strong suppression at small particle distances, which is generated by the short-range repulsive core in the central part of the realistic interaction. The propability density of finding any two nucleons in a nucleus (two-body density) at relative distances smaller than the radius of this core is very small. Furthermore, in addition to the $S$-wave part, the ground state contains a $D$-wave admixture generated by the strong and long-ranged tensor potential. The $D$-wave component and thus the tensor force are essential for nuclear binding, not only in the case of the deuteron. A detailed illustration on these correlations is given for example in Refs. \cite{FeNe98,NeFe03,RoNe04,RoHe05}.  

The central step on the way towards nuclear structure calculations for heavy nuclei based on realistic potentials is the combination of tractable many-body approximations with an appropriate description of interaction-induced correlations. In most cases, this is effectively achieved by converting the bare realistic interaction into an effective interaction adapted to the available model space. In addition to traditional methods, like the Brueckner $G$-matrix \cite{Day67}, several new approaches have been developed recently,  e.g., the $V_{\text{low}k}$ renormalization group method \cite{BoKu03,BoKu03b}. 

We are going to treat the strong short-range correlations in the framework of the unitary correlation operator method (UCOM) \cite{FeNe98,NeFe03,RoNe04,RoHe05}. This approach offers two complementary but equivalent views on correlations: The short-range central and tensor correlations are imprinted into uncorrelated many-body states by a unitary transformation. The unitary operator of this transformation is constructed in a representation-independent operator form, which comprises the physics of the dominant short-range correlations. Alternatively, the correlation operator can be used to transform the operators of the relevant observables. From the transformation of the Hamiltonian including a realistic NN interaction, we obtain a correlated interaction $\VO_{\UCOM}$ which can be used in conjunction with uncorrelated many-body states. This correlated interaction was employed successfully in different many-body methods for light and medium-mass nuclei \cite{RoHe05,RoNe04}. 

The aim of this paper is to demonstrate the capabilities of correlated realistic NN-interactions in the description of ground state properties of nuclei across the whole mass range from \elem{He}{4} to \elem{Pb}{208}. We summarize the elements of the unitary correlation operator method in Sec. \ref{sec:ucom} and illustrate the properties of the correlated interaction $\VO_{\UCOM}$. As the simplest many-body approximation we employ the Hartree-Fock scheme. Implementation and results for ground states of closed shell nuclei are discussed in Sec. \ref{sec:hf}. The impact of residual correlations is investigated by means of many-body perturbation theory based on the Hartree-Fock ground state in Sec. \ref{sec:pt}.

\section{The Unitary Correlation Operator Method (UCOM)}
\label{sec:ucom}

\subsection{Unitary correlation operators}

In the framework of the unitary correlation operator method (UCOM) the dominant short-range correlations are described by a unitary transformation with a correlation operator $\CO$. The explicit operator form of the unitary correlator is constructed following the physical mechanism by which the realistic NN-interactions induce correlations into the many-body state, as discussed in detail in Refs. \cite{RoHe05,RoNe04,NeFe03,FeNe98}. This distinguishes UCOM from other approaches employing unitary transformations to describe correlations, like the Lee-Suzuki transformation \cite{SuLe80,CaNa02,NaOr02} or the unitary model operator approach \cite{FuOk04,SuOk94,ShWa67}, which are entirely formulated in terms of matrix elements. 

The correlation operator is expressed as a product of two independent unitary operators $\CO_{\Omega}$ and $\CO_{r}$ describing short-range tensor and central correlations, respectively:
\eq{ \label{eq:correlator}
  \CO = \CO_{\Omega} \CO_{r}
  = \exp\!\Big[-\ii \sum_{i<j} \gO_{\Omega,ij} \Big]\;
    \exp\!\Big[-\ii \sum_{i<j} \gO_{r,ij} \Big]\;.
}
Each of them is written as an exponential of Hermitian two-body generators $\gO_{\Omega}$ and $\gO_{r}$. 

The generator $\gO_{r}$ in the central correlation operator shall describe the correlations induced by the repulsive core of the interaction---it has to shift close-lying nucleons apart. Formally, this is achieved by a radial distance-dependent shift in the relative coordinate of two particles. Such radial shifts are generated by the projection of the relative momentum $\qOV  = \frac{1}{2} [\pOV_1 - \pOV_2]$ onto the distance vector $\rOV = \xOV_1 - \xOV_2$ of two particles:
\eq{
  \qO_r 
  = \frac{1}{2} \big[\tfrac{\rOV}{\rO}\cdot\qOV 
    + \qOV\cdot\tfrac{\rOV}{\rO} \big] \;.
}
The distance-dependence is described by a function $s_{ST}(r)$ for each spin-isospin channel, leading to 
\eq{ \label{eq:central_generator}
  \gO_r 
  = \sum_{S,T} \frac{1}{2} 
    [ s_{ST}(\rO)\, \qO_r + \qO_r\, s_{ST}(\rO) ]\, \PiO_{ST}\;,
}
where $\PiO_{ST}$ is the projection operator onto two-body spin $S$ and isospin $T$. 

The generator $\gO_{\Omega}$ has to describe the characteristic entanglement between spin and spatial orientation of two nucleons induced by the tensor force. It has the structure of a tensor operator,
\eqmulti{
  \tensorRQO
  =& \tfrac{3}{2} \big[(\sigmaOV_1\!\cdot\rOV)(\sigmaOV_2\!\cdot\qOV_{\Omega}) 
     + (\sigmaOV_1\!\cdot\qOV_{\Omega})(\sigmaOV_2\!\cdot\rOV)\big] \;,
}
where in comparison to the usual tensor operator of the interaction one of the distance operators is replaced by the tangential component of the relative momentum operator:
\eq{
  \qOV_{\Omega} 
  = \qOV - \frac{\rOV}{\rO} \qO_r
  = \frac{1}{2\rO^2}(\LOV\times\rOV - \rOV\times\LOV) \;.
}
Supplemented by a function $\vartheta_T(r)$ describing the strength of the tensor correlations as a function of the inter-particle distance, this defines the generator 
\eq{ \label{eq:tensor_generator}
  \gO_{\Omega} 
  = \sum_{T} \vartheta_T(r)\; \tensorRQO\; \PiO_{1T} \;
}
which acts only in the spin $S=1$ subspace. The effect of this generator is best illustrated by considering correlated states.

\subsection{Correlated states}

If we apply the unitary correlation operator $\CO$ to an uncorrelated many-body state $\ket{\Psi}$, a new correlated many-body state 
\eq{ \label{eq:corr_state}
  \ket{\corr{\Psi}} = \CO\; \ket{\Psi} 
}
results. In the simplest case, for example in a Hartree-Fock calculation, the uncorrelated state is a Slater determinant. The unitary transformation, however, maps it onto a correlated state, which includes the dominant short-range correlations and cannot be represented by a single or a few Slater determinants anymore. 

In two-body space, the analytic form of correlated states can be worked out easily \cite{FeNe98,NeFe03,RoNe04,RoHe05}. For simplicity, we assume $LS$-coupled two-body states of the structure
\eq{
  \ket{\Psi} 
  = \ket{\Phi_{\text{cm}}} \otimes \ket{\phi (L S) J T} \;,
}
where $M$ and $M_T$ are omitted for brevity. The correlation operators do not act on the center of mass component $\ket{\Phi_{\text{cm}}}$ of the two-body state, only the relative part is transformed. In coordinate representation, the relative two-body wave function resulting from the transformation with the central correlator $\cO_{r}=\exp(-\ii \gO_r)$ 
\footnote{Correlation operators in two-body space are denoted by small letters, those in a general $A$-body space by capital letters. The same convention applies to other operators.}
reads
\eqmulti{
  &\matrixe{\rV}{\cO_r}{\phi (L S) J T} =\\
  &\quad= \frac{\Rm(r)}{r}\sqrt{\smash\DRm(r)}\;\; 
    \braket{\Rm(r)\tfrac{\rV}{r}}{\phi (L S) J T} \;.
}
This corresponds to a norm-conserving coordinate transformation $r \mapsto \Rm(r)$. The transformation with the Hermitian adjoint correlator $\ccO_{r}$ leads to an analogous expression with the inverse correlation function $\Rp(r)$, where $\Rpm[\Rmp(r)]=r$. The correlation functions $\Rpm(r)$ are related to the shift function $s(r)$ in \eqref{eq:central_generator} by 
\eq{
  \int_{r}^{\Rpm(r)} \frac{\dd{\xi}}{s(\xi)} = \pm 1 \;,
}
where spin and isospin indices have been omitted for brevity. 

The action of the tensor correlation operator $\cO_{\Omega}=\exp(-\ii \gO_{\Omega})$ onto $LS$-coupled two-body states can be evaluated directly using matrix elements of the tensor operator \tensorRQO\ contained in the generator \cite{NeFe03}. Two-body states with $L=J$ are invariant under transformation with the tensor correlator: 
\eq{ \label{eq:corr_tensor_states1}
  \cO_{\Omega} \ket{\phi (J S) J T}
  = \ket{\phi (J S) J T} \;.
}
Only states with $L=J\pm1$ are affected by $\cO_{\Omega}$ and transform like
\eqmulti{ \label{eq:corr_tensor_states2}
  \cO_{\Omega} \ket{\phi (J\pm 1,1) J T}
  &= \cos\theta_J(\rO)\, \ket{\phi (J\pm 1,1) J T} \\
  &\mp\, \sin\theta_J(\rO)\, \ket{\phi (J\mp 1,1) J T}
}
with $\theta_J(\rO) = 3 \sqrt{J(J+1)}\; \vartheta(\rO)$. The tensor correlation operator thus generates components with $\Delta L=\pm 2$ in the correlated state. If we start with an uncorrelated state with $L=0$, $S=1$, and $J=1$, then the correlated state acquires an additional $L=2$ admixture, whose radial dependence is determined by the tensor correlation function $\vartheta(r)$. 
 
The relations for the correlated two-body states form the basis for the evaluation of the matrix elements of correlated operators without approximations.

\subsection{Correlated operators}

One of the virtues of the description of correlations by a state-independent unitary transformation is that instead of working with correlated states, one can also apply the unitary correlator onto the operators of interest and define correlated operators
\eq{ \label{eq:corr_operator}
  \corr{\AO} = \CCO \AO \,\CO \;.
}
For the calculation of observables, e.g., expectation values or matrix elements, the formulations in terms of correlated operators and correlated states are fully equivalent and one can choose whichever is technically more convenient. Note that when the notion of correlated operators is used, all operators of interest have to be transformed consistently.

The correlated operator $\corr{\AO}$ contains irreducible contributions to all particle numbers, 
\eq{ \label{eq:corr_clusterexp}
  \corr{\AO}
  = \CCO \AO \CO
  = \corr{\AO}^{[1]} + \corr{\AO}^{[2]} + \corr{\AO}^{[3]} + \cdots \;,
}
where $\corr{\AO}^{[n]}$ denotes the irreducible $n$-body part \cite{FeNe98}. Hence, the unitary transformation of a two-body operator --- the NN-interaction for example --- yields a correlated operator containing a two-body contribution, a three-body term, etc. The contributions of terms beyond the two-body order of this cluster expansion depend on the range of the correlators. For correlation functions $s(r)$ and $\vartheta(r)$ of sufficiently short range, three-body and higher order contributions can be neglected \cite{RoNe04,RoHe05}. This defines the two-body approximation of a correlated operator, $\corr{\AO}^{C2} = \corr{\AO}^{[1]} + \corr{\AO}^{[2]}$.

For a Hamiltonian $\HO$ consisting of one-body kinetic energy $\TO=\sum_i \pOV_i^2/(2 m_N)$ and a two-body NN-interaction $\VO=\sum_{i<j}\vO_{ij}$, the correlated operator in two-body approximation reads
\eq{ \label{eq:corr_hamiltonianC2}
  \corr{\HO}^{C2} 
  = \corr{\TO}^{[1]} + \corr{\TO}^{[2]} + \corr{\VO}^{[2]}
  = \TO + \VO_{\UCOM} \;.
}
The one-body contribution to the correlated Hamiltonian is just the uncorrelated kinetic energy. The two-body part consists of a contribution of the correlated kinetic energy $\corr{\TO}^{[2]}$ and the correlated potential $\corr{\VO}^{[2]}$. Together these two-body contributions define a correlated or  effective two-body interaction $\VO_{\UCOM}=\sum_{i<j}\vO_{\UCOM,ij}$. The unitary transformation preserves the symmetries of the bare operators. Therefore, the correlated interaction has the same symmetries as the underlying NN potential, i.e. translational, rotational, Galilei, and parity invariance.

Two inherent properties of $\VO_{\UCOM}$ are of great importance for the practical application: 
(\emph{i}) The correlated interaction  $\VO_{\UCOM}$ is phase-shift equivalent to the bare potential one starts with. Hence the defining property of modern realistic NN-interactions --- the reproduction of experimental phase shifts from nucleon-nucleon scattering with high precision --- is preserved and $\VO_{\UCOM}$ can be considered a realistic potential in its own right. 

(\emph{ii}) Based on the correlation operator \eqref{eq:correlator} an explicit operator form of the correlated interaction $\VO_{\UCOM}$ can be derived. The details of this derivation and the structure of the resulting interaction operators are discussed in Refs. \cite{RoNe04,NeFe03,FeNe98}. This property distinguishes UCOM from other approaches to derive phase-shift equivalent effective interactions, like the $V_{\text{low}k}$ approach \cite{BoKu03,BoKu03b}, which is formulated entirely on the level of matrix elements. In many-body schemes which do not allow the use of partial-wave matrix elements, like the Fermionic Molecular Dynamics approach \cite{Feld90,FeSc00,RoNe04}, the knowledge of a closed operator form of the effective interaction is indispensable.  

Besides the Hamiltonian all other observables like radii or transition strengths can and must be correlated in the same way to be consistent. In most other schemes to derive effective interactions it is very difficult or not even obvious how to derive consistently the corresponding effective observables.

\subsection{Correlated matrix elements}
\label{sec:ucom_me}

For the use of the correlated interaction $\VO_{\UCOM}$ in standard many-body schemes based on an orthogonal single-particle basis we have to evaluate appropriate two-body matrix elements. Let us assume a spherical harmonic-oscillator basis as it will be used in the Hartree-Fock calculations discussed in Sec. \ref{sec:hf}. 

In a first step we consider $LS$-coupled harmonic-oscillator two-body matrix elements of the form
\eqmulti{ \label{eq:corr_me_VUCOM}
  &\matrixe{n(LS)J T}{\vO_{\UCOM}}{n'(L'S)J T} \\
  &\quad= \matrixe{n(LS)J T}{\ccO_{r}\ccO_{\Omega}\,\HO\,\cO_{\Omega}\cO_{r} 
    - \TO}{n'(L'S)J T} \;,
}
where $n,n'=0,1,2,...$ are the radial oscillator quantum numbers of the relative two-body states ($M$ and $M_T$ are omitted). One can, of course, use the operator representation of $\vO_{\UCOM}$ and compute these two-body matrix elements directly. However, it is more convenient to map the correlation operators back onto the $LS$-coupled two-body states and thus compute the correlated matrix elements using uncorrelated operators and correlated states \cite{RoHe05}. For the tensor correlations this is a substantial simplification, since the tensor correlated two-body states \eqref{eq:corr_tensor_states1} and \eqref{eq:corr_tensor_states2} are simple in comparison to the corresponding tensor correlated Hamilton operator. 

In Ref. \cite{RoHe05} we have developed a hybrid scheme with the central correlator $\cO_{r}$ applied to the operators and the tensor correlator $\cO_{\Omega}$ applied to the two-body states. From the computational point of view, this turns out to be the most efficient approach and will be the basis for the following numerical calculations. For the sake of brevity, we do not repeat the relevant expressions here.

In the second step we have to transform the $LS$-coupled relative matrix elements of $\vO_{\UCOM}$ into matrix elements with respect to antisymmetrized $jj$-coupled two-body states $\ket{n_1 l_1 j_1, n_2 l_2 j_2; J T}$. For the harmonic oscillator basis this is achieved by the well-known Talmi-Moshinsky transformation \cite{Talm52,Mosh59}. Including angular momentum recoupling one obtains the following relation: 
\eqmulti{
  &\matrixe{n_1 l_1 j_1, n_2 l_2 j_2; J T}
     {\vO_{\UCOM}}{n'_1 l'_1 j'_1, n'_2 l'_2 j'_2; J T} = 
    \\[3pt]
  &\quad= \sqrt{[j_1] [j_2] [j'_1] [j'_2]}
     \sum_{L,L',S} \sum_{N, \Lambda} \sum_{\nu,\lambda}
     \sum_{\nu',\lambda'}\sum_{j} \\[3pt]
  &\quad\times \ninejsymb{l_1}{l_2}{L}{\half}{\half}{S}{j_1}{j_2}{J}
    \ninejsymb{l'_1}{l'_2}{L'}{\half}{\half}{S}{j'_1}{j'_2}{J}
    \sixjsymb{\Lambda}{\lambda}{L}{S}{J}{j}
    \sixjsymb{\Lambda}{\lambda'}{L'}{S}{J}{j}
    \\[3pt]
  &\quad\times\talmimosh{N}{\Lambda}{\nu}{\lambda}{n_1}{l_1}{n_2}{l_2}{L}\; 
    \talmimosh{N}{\Lambda}{\nu'}{\lambda'}{n'_1}{l'_1}{n'_2}{l'_2}{L'}
    \\[3pt]
  &\quad \times [j][S][L][L']\; (-1)^{L+L'}\; \{1- (-1)^{\lambda+S+T}\} \\[3pt]
  &\quad \times \matrixe{\nu (\lambda S) j T}{\vO_{\UCOM}}{\nu' (\lambda' S) j T} \;,
}
where $[j] \equiv 2j+1$. In addition to $9j$ and $6j$ symbols, the harmonic oscillator brackets $\langle\!\langle...|...\rangle\!\rangle$ appear \cite{KaKa01,BuMe96}. Three of the above summations can be eliminated right away. For given $N$, $\Lambda$, $\lambda$, and $\lambda'$, the possible values of $\nu$ and $\nu'$ can be determined directly from the relation $(2N + \Lambda) + (2\nu + \lambda) = (2n_1 + l_1) + (2 n_2 +l_2)$. The factor $(1- (-1)^{\lambda+S+T})$ resulting from the antisymmetrization removes all terms with even values of $\lambda+S+T$ and can be used to eliminate the $S$ summation for given $\lambda$ and $T$. 

Of course, this procedure is not restricted to the interaction matrix elements. We evaluate other correlated observables, e.g. correlated rms-radii, in an analogous way.

\subsection{Optimal correlation functions}
\label{sec:ucom_optcorr}

Given the formal expressions for the matrix elements of the correlated interaction, the only remaining task is to determine the optimal correlation functions $\Rp(r)$ and $\vartheta(r)$ for the realistic NN potential under consideration. In this paper we will restrict ourselves to the Argonne V18 (AV18) potential \cite{WiSt95}. The determination of the optimal correlation functions was discussed in \cite{RoHe05} and we only summarize the important results here.

The easiest way to determine optimal correlation functions is a variational calculation in the two-body system. For each combination of two-body spin $S$ and isospin $T$ we minimize the expectation value of the correlated Hamiltonian by varying the central and tensor correlation functions. To this end, the following parameterizations for the central correlation functions are used for the even and odd channels, respectively:
\eqmulti{
  \Rp^{\text{I}}(r) 
  &= r + \alpha\, (r/\beta)^{\eta} \exp[-\exp(r/\beta)]  \;,\\
  \Rp^{\text{II}}(r)
  &= r + \alpha\, [1 - \exp(-r/\gamma)] \exp[-\exp(r/\beta)] \;.
}
For the tensor correlation functions the following form turns out to be most suitable:
\eq{
  \vartheta(r) 
  = \alpha\, [1 - \exp(-r/\gamma)] \exp[-\exp(r/\beta)] \;.
}
The optimal parameter values as determined in Ref. \cite{RoHe05} are summarized in Tables \ref{tab:corr_centralpara} and \ref{tab:corr_tensorpara}.
 
\begin{table}
\begin{ruledtabular}
\begin{tabular}{c c c c c c c}
$S$ & $T$ & Param. &  $\alpha$ [fm] & $\beta$ [fm] & $\gamma$ [fm] & $\eta$ \\
\hline
0 & 0 & II &  0.7971 &  1.2638  &  0.4621  &  ---     \\
0 & 1 & I  &  1.3793 &  0.8853  &  ---     &  0.3724 \\
1 & 0 & I  &  1.3265 &  0.8342  &  ---     &  0.4471 \\
1 & 1 & II &  0.5665 &  1.3888  &  0.1786  &  ---     \\
\end{tabular}
\end{ruledtabular}
\caption{Parameters of the central correlation functions $\Rp(r)$ for
the AV18 potential obtained from two-body energy minimization.}
\label{tab:corr_centralpara}
\end{table}
\begin{table}
\begin{ruledtabular}
\begin{tabular}{c c c c c c}
$S$ & $T$ & $I_{\vartheta}$ [fm${}^3$] &  $\alpha$ & $\beta$ [fm] & $\gamma$ [fm] \\
\hline
1 & 0 & 0.08 &  541.29  &  1.2215  &  1000.0  \\
1 & 0 & 0.09 &  536.67  &  1.2608  &  1000.0  \\
1 & 0 & 0.10 &  531.03  &  1.2978  &  1000.0  \\
\end{tabular}
\end{ruledtabular}
\caption{Parameters of the triplet-even tensor correlation function $\vartheta(r)$ for
the AV18 potential with different values $I_{\vartheta}$ for the
range constraint.}
\label{tab:corr_tensorpara}
\end{table}

The tensor correlation function for $S=1$ and $T=1$ (triplet-odd channel) turns out to be at least one order of magnitude smaller than for $S=1$ and $T=0$ (triplet-even channel) \cite{RoHe05}. This is a consequence of the much weaker tensor potential in this channel. In order to simplify the present study, we will not include any tensor correlator in the triplet-odd channel and concentrate on the impact of the dominant tensor correlations in the triplet-even channel.

A crucial point is the range of the tensor correlations. The tensor force in the triplet-even channel is very long-ranged due to its origin from one-pion exchange. In an isolated two-body system, i.e. the deuteron, the associated tensor correlations will be present up to large interparticle distances. The ramification of this is the long-range $D$-wave admixture in the deuteron wave function. Therefore, we will obtain a long-range tensor correlation function if we employ an unconstrained energy minimization in the two-body system.
 
In the many-body system, the tensor interaction with other nucleons will prevent the formation of the long-range component of tensor correlations between a pair of nucleons. Effectively, the long-range tensor correlations are screened. We anticipate this many-body screening effect by imposing a constraint on the range of the tensor correlator defined by the volume integral
\eq{ \label{eq:tensor_constraint}
  I_{\vartheta}
  = \int \dd{r}\, r^2\; \vartheta(r) \;.
}
The results of the constrained minimization in the two-body system for different values of the measure $I_{\vartheta}$ are summarized in Table \ref{tab:corr_tensorpara}. 

The restriction to short-range tensor correlators is helpful also in connection to the two-body approximation for correlated operators. If a long-range tensor correlator were used, like suggested by the deuteron wave function, then the higher-orders of the cluster expansion would yield sizeable and nontrivial contributions. In fact, they represent the aforementioned many-body screening of long-range tensor correlations. By restricting the range of the tensor correlators these higher-order contributions are reduced from the outset. 

The choice of an appropriate value of $I_{\vartheta}$ requires information beyond the two-body problem. All other parameters are fixed on the level of the two-nucleon system alone. One strategy to fix $I_{\vartheta}$ is by means of an exact few-body calculation using $\VO_{\UCOM}$. As we have shown in Ref. \cite{RoHe05}, the exact binding energies of $\elem{H}{3}$ and $\elem{He}{4}$ obtained in no-core shell model calculations for different $I_{\vartheta}$ map out the Tjon-line . Moreover, for $I_{\vartheta}\approx0.09\,\text{fm}^3$, the exact calculation based on $\VO_{\UCOM}$ reproduces the experimental binding energies for $A\leq4$ quite well. The fact that the experimental energies are matched without including a genuine three-body force and the induced three-body contributions of the cluster expansion indicates that the net effect of those missing three-body terms on the ground state energies vanishes. In other words, the three-body contribution of the cluster expansion cancels the genuine three-body force \cite{RoHe05}.

We will use the triplet-even tensor correlator for $I_{\vartheta}=0.09\,\text{fm}^3$ as the optimal correlator for the present study of heavier nuclei. This fixes the correlated interaction and all the following calculations are therefore completely parameter-free.

\subsection{Illustration}
\label{sec:ucom_illu}

As a first demonstration of the impact of the unitary transformation, we perform a naive shell-model-type calculation of the ground state energy for various nuclei. We assume an uncorrelated many-body state given by a single Slater determinant built of harmonic oscillator single-particle states. Clearly, this independent-particle state does not contain any of the relevant many-body correlations. We successively apply the central and the tensor correlation operators, $\CO_r$ and $\CO_{\Omega}$, respectively, and investigate their effect on the energy expectation value for a many-body Hamiltonian containing the AV18 potential. As discussed earlier, we map the correlation operators onto the Hamiltonian and employ the two-body approximation, which leads to the correlated interaction $\VO_{\UCOM}$. Using the correlated two-body matrix elements constructed in Sec. \ref{sec:ucom_me}, we can directly evaluate the expectation value of the correlated Hamiltonian. 

\begin{figure}
\begin{center}
\includegraphics[width=1\columnwidth]{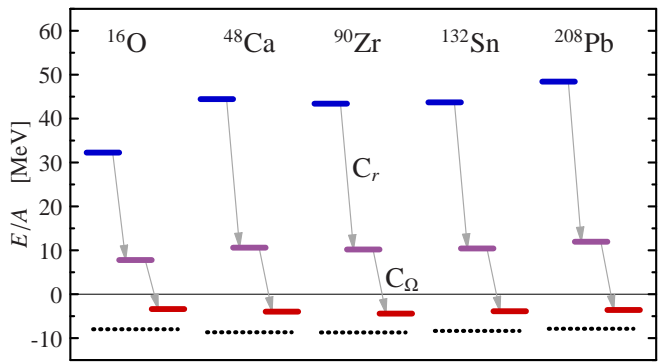}
\end{center}
\caption{(Color online) Effect of the unitary transformation on energy expectation values of different nuclei with simple shell-model Slater determinants. The four levels for each nucleus indicate (from top to bottom) the expectation value of the bare Hamiltonian, the centrally correlated, the fully correlated Hamiltonian, and the experimental binding energy per particle, respectively. The AV18 potential with the optimal correlators for $I_{\vartheta}=0.09\,\text{fm}^3$ is used.}
\label{fig:hf_levelillu}
\end{figure}

In Fig. \ref{fig:hf_levelillu} the expectation values of the uncorrelated, the central correlated, and the fully correlated Hamiltonian are displayed together with the experimental ground state energies for various nuclei ranging from \elem{O}{16} to \elem{Pb}{208}. The oscillator parameter is chosen such that the expectation value of the fully correlated Hamiltonian is minimized for the nucleus under consideration. The center of mass kinetic energy is subtracted.  

Evidently, the expectation value of the bare Hamiltonian with an uncorrelated Slater determinant is positive and large---all nuclei are unbound. The proper inclusion of correlations is crucial for obtaining bound nuclei. By invoking the central correlation operator $\CO_r$, i.e., by including the correlations induced by the repulsive core of the interaction, the energy is reduced significantly. However, the inclusion of these central correlations alone is not sufficient to obtain self-bound nuclei. Employing the tensor correlation operator $\CO_{\Omega}$ in addition, i.e., accounting for the short-range part of the correlations caused by the tensor component of the interaction, the energy is lowered further, and we eventually obtain bound nuclei troughout the whole mass range. This emphasizes the importance of the tensor part of realistic NN-interactions and of the associated correlations in the nuclear many-body problem.

This simplistic calculation highlights two important points: (\emph{i}) The unitary correlation operators provide a very efficient means of describing the state-independent short-range correlations induced by the repulsive core and tensor part of the potential. Throughout the nuclear chart, one obtains bound nuclei starting from a simple Slater determinant as an uncorrelated many-body state. In comparison to the uncorrelated expectation value, the correlators reduce the ground state energy  by typically more than 40 MeV per nucleon for the AV18 potential (see Fig. \ref{fig:hf_levelillu}). Motivated by this observation we use the correlated realistic interaction for Hartree-Fock calculations as described in Sec. \ref{sec:hf}.

(\emph{ii}) The resulting binding energy is typically smaller than the experimental binding energy. This indicates that residual long-range correlations not accounted for by the explicit unitary transformations have to be considered. This is fully in-line with the results from no-core shell-model calculations using $\VO_{\UCOM}$ discussed in Ref. \cite{RoHe05}. Those missing state-dependent correlations need to be included through the degrees of freedom of the available many-body states. It is remarkable that the deviation from the experimental binding energy per nucleon is practically constant over the whole mass range. This already hints that the deviation is not dominated by missing three-body forces. We will discuss this point in detail in Sec. \ref{sec:pt}.

\section{UCOM Hartree-Fock Scheme}
\label{sec:hf}

Using the correlated realistic NN-interaction we set up a Hartree-Fock scheme and investigate the behavior of the ground state solutions across the nuclear chart. Furtheron, the Hartree-Fock solutions form the basis for an improved treatment of the nuclear many-body problem.

\subsection{Formulation}

According to the basic assumption of the Hartree-Fock scheme, the many-body state is approximated by a single Slater determinant
\eq{ \label{eq:hf_hfstate}
  \ket{\text{HF}} 
  = \AC\; (\ket{\alpha_1}\otimes\ket{\alpha_2}\otimes\cdots\otimes\ket{\alpha_A}),
}
where $\AC$ is the antisymmetrization operator acting on an $A$-body product state. The single-particle states $\ket{\alpha_i}$ are used as variational degrees of freedom in a minimization of the expectation value of the many-body Hamiltonian. The formal variational solution of the many-body problem using the trial state \eqref{eq:hf_hfstate} leads to the well known Hartree-Fock equations \cite{RiSc80}.

As illustrated in Sec. \ref{sec:ucom_illu}, the explicit inclusion of correlations beyond the independent-particle state $\ket{\text{HF}}$ is crucial when starting from realistic NN-interactions like the AV18 potential. By applying the unitary correlation operator $\CO$ to $\ket{\text{HF}}$, we obtain a correlated many-body state which has the dominant correlations imprinted. By switching from the picture of correlated states to the picture of correlated operators and invoking the two-body approximation, we formally recover a standard Hartree-Fock problem. However, the Hamiltonian entering into the Hartree-Fock calculation now consists of the kinetic energy and the correlated interaction $\VO_{\UCOM}$. In order to account for the center of mass contribution to the energy, we subtract the operator $\TO_{\cm}$ of the center of mass kinetic energy. This leads to the correlated  intrinsic Hamiltonian $\corr{\HO}_{\intr}$ in two-body approximation,         
\eq{
  \corr{\HO}_{\intr} 
  = \TO - \TO_{\cm} + \VO_{\UCOM} 
  = \TO_{\intr} + \VO_{\UCOM} \;,
}
where the superscript ``$C2$'' indicating the two-body approximation has been omitted [cf. Eq. \eqref{eq:corr_hamiltonianC2}]. The correlated Coulomb interaction together with charge dependent terms of the NN-potential are included in $\VO_{\UCOM}$ and are not written separately. The intrinsic kinetic energy operator can  be expressed in terms of the relative two-body momentum operator $\qOV$ alone:
\eq{
  \TO_{\intr} 
  = \TO - \TO_{\cm} 
  = \frac{2}{A} \frac{1}{m_N} \sum_{i<j}^{A} \qOV_{ij}^2 \;,
}
where we have assumed equal proton and neutron masses and thus a reduced mass $\mu=m_N/2$. Thus, $\corr{\HO}_{\intr}$ technically has the structure of a pure two-body operator which facilitates the implementation. Although this explicitly Galilei-invariant Hamiltonian does not exclude the possibility of center-of-mass excitations, their contribution to the ground state is expected to be small. A stringent but computationally expensive approach requires an explicit center-of-mass projection \cite{Schm02,RoSc04}.

We formulate the Hartree-Fock scheme in a basis representation using harmonic oscillator states in order to use the correlated matrix elements of realistic NN-potentials discussed in Sec. \ref{sec:ucom_me}. The Hartree-Fock single particle states $\ket{\alpha}$ are written as
\footnote{$\alpha = \{\nu l j m m_t\}$ is used as a collective index for all quantum numbers of the HF single particle states.} 
\eqmulti{
  \ket{\alpha} 
  = \ket{\nu l j m m_t} 
  = \sum_n C^{(\nu l j m m_t)}_{n} \ket{n l j m m_t} \;,
}
where $\ket{n l j m m_t}$ denotes a harmonic oscillator eigenstate with radial quantum number $n$, orbital angular momentum $l$, total angular momentum $j$ with projection $m$, and isospin projection quantum number $m_t$. Assuming spherical symmetry, only oscillator states with the same quantum numbers $l$, $j$, and $m$ can contribute in the expansion. In the following, we will restrict ourselves to constrained or closed-shell calculations, where $C^{(\nu l j m m_t)}_{n}=C^{(\nu l j m_t)}_{n}$ is independent of $m$.

The expansion coefficients $C^{(\nu l j m_t)}_{n}$ are used as variational parameters for the minimization of the energy expectation value. The formal variation leads to a non-linear matrix eigenvalue problem determining the optimal coefficients \cite{RiSc80}:
\eq{ \label{eq:hf_hfeigenproblem}
  \sum_{\bar{n}} h^{(l j m_t)}_{n\bar{n}} C^{(\nu l j m_t)}_{\bar{n}} 
  = \epsilon^{(\nu l j m_t)} C^{(\nu l j m_t)}_{n} \;,
}
where $\epsilon^{(\nu l j m_t)}$ are the corresponding single-particle energy eigenvalues.  The matrix elements of the HF single-particle Hamiltonian $h^{(l j m_t)}_{n\bar{n}}$ are given by 
\eq{ \label{eq:hf_hfhamiltonian}
  h^{(l j m_t)}_{n\bar{n}}
  = \sum_{l',j',m'_t} \sum_{n',\bar{n}'} 
    H^{(l j m_t l' j' m'_t)}_{n n'; \bar{n} \bar{n}'}
    \varrho^{(l' j' m'_t)}_{n' \bar{n}'} 
}
with the one-body density matrix 
\eq{ \label{eq:hf_hfdensitymat}
  \varrho^{(l j m_t)}_{n \bar{n}}
  = \sum_{\nu} O^{(\nu l j m_t)}
    C^{(\nu l j m_t)^{\star}}_{\bar{n}} C^{(\nu l j m_t)}_{n} \;.
}
Here $O^{(\nu l j m_t)}$ is the number of occupied magnetic sublevels in the respective shell, which is simply $O^{(\nu l j m_t)}=2j+1$ for closed shell configurations. Via the density matrix, the single-particle Hamilton matrix itself depends on the coefficients $C^{(\nu l j m_t)}_{n}$, entailing the non-linear character of the eigenvalue problem \eqref{eq:hf_hfeigenproblem}. 

The essential ingredient for the single-particle Hamilton matrix \eqref{eq:hf_hfhamiltonian} are the $m$-averaged antisymmetric two-body matrix elements of the correlated intrinsic Hamiltonian:     
\eqmulti{ \label{eq:hf_twobodyme}
  &H^{(l j m_t l' j' m'_t)}_{n n', \bar{n} \bar{n}'} \\
  &\quad= \frac{1}{(2j+1)(2j'+1)} \sum_{m,m'} \\
  &\quad\times \matrixe{n l j m m_t, n' l' j' m' m'_t}
    {\corr{\HO}_{\intr}}{\bar{n} l j m m_t, \bar{n}' l' j' m' m'_t} \;.
}
Instead of starting from uncoupled two-body matrix elements, we can employ $jj$-coupled two-body matrix elements and cast Eq. \eqref{eq:hf_twobodyme} into the more convenient form
\eqmulti{ \label{eq:hf_twobodymejj}
  &H^{(l j m_t l' j' m'_t)}_{n n'; \bar{n} \bar{n}'} \\
  &\quad= \sum_{J,T,M_T} \frac{(2J+1)}{(2j+1)(2j'+1)} 
    \braket{\half m_t; \half m'_t}{T M_T}^2 \\
  &\quad\times \matrixe{n l j, n' l' j'; J T M_T}
    {\corr{\HO}_{\intr}}{\bar{n} l j, \bar{n}' l' j'; J T M_T} \;. 
}
Since the intrinsic Hamiltonian includes the Coulomb potential as well as charge dependent terms of the correlated NN interaction, we explicitly indicate the $M_T$-dependence of the matrix elements.

\subsection{Implementation \& convergence}

The implementation of the Hartree-Fock procedure using correlated matrix elements of realistic NN-potentials is straightforward. The harmonic oscillator basis is truncated in the major oscillator quantum number $e=2n+l \leq e_{\max}$. Additional truncations with respect to the radial quantum number $n$ or the orbital angular momentum $l$ can be employed. The optimal oscillator parameter for a given nucleus is determined by an explicit minimization within a set of oscillator parameters.

The major computational effort goes into the calculation of the correlated $jj$-coupled matrix elements of the interaction. They are computed separately and stored to disk for given $e_{\max}$ (or $n_{\max}$, $l_{\max}$) and oscillator length $a_{\text{HO}}$. Once calculated, they are used as input for the Hartree-Fock calculations as well as for many-body perturbation theory, etc. The two-body matrix elements for the intrinsic kinetic energy and other observables are handled in the same way. Thus, the conceptionally and technically demanding step of computing correlated matrix elements of various operators is completely separated from the simple task of solving the nonlinear single-particle eigenvalue problem \eqref{eq:hf_hfeigenproblem}. The latter is solved in a standard iterative procedure until full self-consistency is obtained.

\begin{figure}
\begin{center}
\includegraphics[width=0.75\columnwidth]{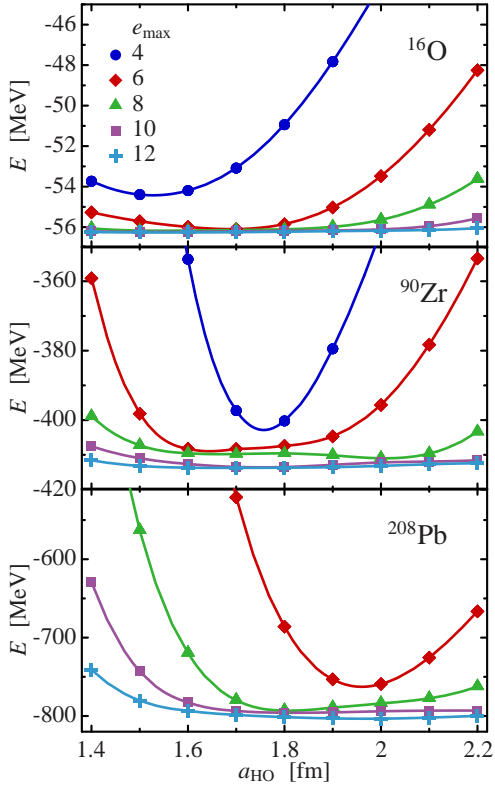}
\end{center}
\caption{(Color online) Ground state energy of \elem{O}{16}, \elem{Zr}{90}, and \elem{Pb}{208} as a function of the oscillator length $a_{\text{HO}}$ for different basis sizes $e_{\max}$. The correlated AV18 potential with $I_{\vartheta}=0.09\,\text{fm}^3$ is used. 
}
\label{fig:hf_convergence}
\end{figure}

Besides the ground state energy, we consider root-mean-square (rms) radii as a second simple observable. As for all observables, the unitary transformation has to be applied consistently to the corresponding operators. The translationally invariant form of the operator for the square radius can be written as follows:
\eq{ \label{eq:hf_radius}
  \RO_{\text{sq}} 
  = \frac{1}{A} \sum_{i} (\xOV_i - \XOV_{\cm})^2 
  = \frac{1}{A^2} \sum_{i<j} \rO_{ij}^2 \;,
}
where $\XOV_{\cm} = \frac{1}{A} \sum_i \xOV_i$. The unitary transformation with the central correlator can be evaluated directly on the operator level and generates an additional two-body term,
\eq{
  \corr{\RO}_{\text{sq}}^{C2} 
  = \RO_{\text{sq}} + \frac{1}{2A} \sum_{i<j} [\Rp(\rO_{ij})^2 - \rO_{ij}^2] \;,
}
where we have suppressed the spin-isospin dependence of the correlation function $\Rp(r)$. The tensor correlator does not affect this operator since it only depends on the relative distance. The square-root of the corresponding expectation value for the HF ground state yields the correlated point rms-radius. The proton point rms-radius is obtained in the analogous manner, and, after adding the standard correction for proton and neutron size, leads to the correlated charge radius $R_{\text{ch}}$. We note that the effect of the unitary transformation on the rms-radii is marginal. The difference between the correlated and the uncorrelated rms-radius is very small for all particle numbers (typically between $0.01$ and $0.02\,\text{fm}$). Pictorially, the central correlators modify the wave function only at small interparticle distances, whereas the square radius operator \eqref{eq:hf_radius} is more sensitive to the behavior at large distances.

As a first benchmark we investigate the convergence properties of the UCOM-HF calculations  for different closed-shell nuclei. Figure \ref{fig:hf_convergence} depicts the ground state energy for \elem{O}{16}, \elem{Zr}{90}, and \elem{Pb}{208} as a function of the oscillator length $a_{\text{HO}}$. With increasing size of the single-particle space, characterized by the truncation parameter $e_{\max}$, we observe a very nice convergence for all nuclei. In general, bases with $e_{\max} = 10$ or $12$ are sufficient to obtain binding energies which are fully converged and independent of  $a_{\text{HO}}$ over a wide range. All following calculations are performed for $e_{\max}=12$.

\subsection{Ground-state energies \& charge radii}

We now study the global systematics of binding energies and charge radii of closed-shell nuclei using the correlated AV18 potential. In addition to the behavior as function of the mass number, we investigate the dependence on the range of the triplet-even tensor correlation function, i.e., the value of the correlation measure $I_{\vartheta}$.

\begin{figure}
\begin{center}
\includegraphics[width=\columnwidth]{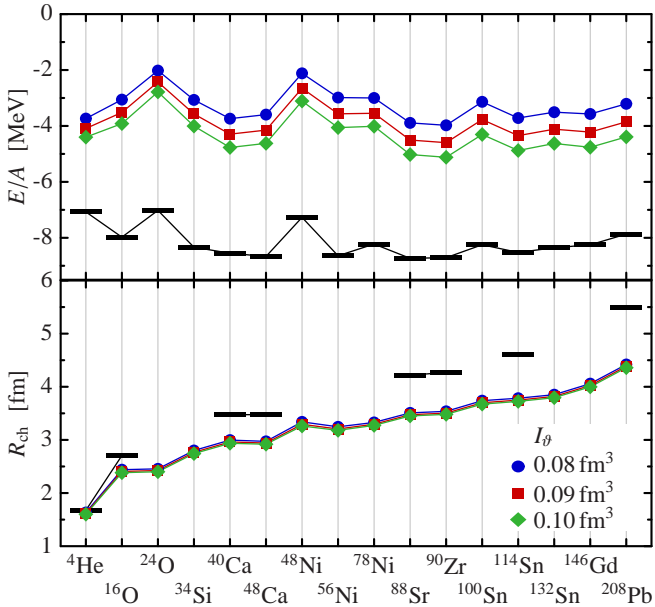}
\end{center}
\caption{(Color online) Ground-state energies and charge radii of selected closed-shell nuclei resulting from a Hartree-Fock calculation with the correlated AV18 potential. Three different ranges for the triplet-even tensor correlator are used: $I_{\vartheta}=0.08\,\text{fm}^3$, $0.09\,\text{fm}^3$, and $0.10\,\text{fm}^3$. The bars indicate  experimental values \cite{AuWa95,VrJa87}. }
\label{fig:hf_stdnucl_EhfRchhf_tensorrange}
\end{figure}

Figure \ref{fig:hf_stdnucl_EhfRchhf_tensorrange} depicts the binding energies and the correlated charge radii obtained from Hartree-Fock calculations for selected closed-shell nuclei ranging from \elem{He}{4} to \elem{Pb}{208}. The different data sets correspond to different values of the integral constraint \eqref{eq:tensor_constraint} on the triplet-even tensor correlator around the optimal value $I_{\vartheta}=0.09\,\text{fm}^3$ (see Sec. \ref{sec:ucom_optcorr}). 

As anticipated from the schematic calculation in Sec. \ref{sec:ucom_illu}, we obtain bound solutions for all nuclei, indicating that the dominant correlations are indeed introduced very efficiently by the unitary correlation operators. In comparison to the experimental data, the binding energies are underestimated and the charge radii are too small. It is remarkable, however, that the systematics of the binding energies is very well reproduced over the full mass range except for an almost constant offset. 

The deviation is not surprising: The no-core shell-model calculations of Ref. \cite{RoHe05} have shown that residual long-range correlations, which are not covered explicitly by the unitary correlation operators, influence the binding energies. Moreover, the three-body contributions of the cluster expansion of the correlated Hamiltonian and the genuine three-body interactions are not included. We have chosen the range of the tensor correlator such that the net three-body contributions to the energy are minimal for systems with $A\le4$. However, it is by no means clear that this remains true for larger systems. Therefore, it is important to disentangle the effects of missing long-range correlations and three-body forces. We will discuss this question in detail in Sec. \ref{sec:pt}, where we assess the impact of long-range correlations in the framework of many-body perturbation theory.

The interplay between missing long-range correlations and three-body terms also determines the behavior of the HF energies as functions of the tensor correlator range, as the  different data sets in Fig. \ref{fig:hf_stdnucl_EhfRchhf_tensorrange} illustrate. The description of longer-ranged tensor correlations is improved by increasing the correlator range, resulting in a lower energy. At the same time, the repulsive three-body contributions of the correlated Hamiltonian, not included in the two-body approximation, become larger. For correlator ranges $I_{\vartheta}\approx 0.09\,\text{fm}^3$, the effect of the induced and the genuine three-body forces on the ground state energy cancels \cite{RoHe05}. Any longer ranged correlator will generate a repulsive net three-body contribution.

\subsection{Single-particle levels \& spin-orbit splittings}
\label{sec:hf_singleparticle}

The energy eigenvalues of the single-particle HF Hamiltonian provide an additional source of information. However, one has to be extremely careful with their interpretation, since they are no direct experimental observables. In the simplest case they can be defined via many-body energy differences of neighboring nuclei.

In a conventional HF treatment without any center of mass correction, the single-particle energy of an occupied state $\ket{\beta}$ corresponds to the change of the energy expectation value $E_A - E_{A-1}(\beta\;\text{removed})$ when removing this state from the $A$-body Slater determinant under the assumption of static single-particle states (Koopmans' theorem). This direct connection does not hold for a HF scheme based on the intrinsic Hamiltonian \cite{KhKa74,JaHa92}. Calculating the change of the energy expectation value when removing one particle from the HF Slater determinant results for $\epsilon_{\beta}<\epsilon_{F}$ in
\eqmulti{
  \epsilon_{\beta}^{\text{corr}} 
  &= E_A - E_{A-1}(\beta\;\text{removed}) = \\
  &= \epsilon_{\beta} - \frac{\expect{\TO_{\intr}}}{A-1} 
    + \frac{2}{m A(A-1)} \sum_{\alpha}^{<\epsilon_F}  
     \matrixe{\alpha \beta}{\qOV^2}{\alpha \beta} \;,
}
where $\expect{\TO_{\intr}}$ is the expectation value of the intrinsic kinetic energy operator and the summations extend over all occupied levels, both with respect to the $A$-body system. The corresponding energy difference for a system with one additional particle in state $\ket{\beta}$ reads for $\epsilon_{\beta}>\epsilon_{F}$ 
\eqmulti{
  \epsilon_{\beta}^{\text{corr}} 
  &= E_{A+1}(\beta\;\text{added}) - E_A = \\
  &= \epsilon_{\beta} 
    - \frac{\expect{\TO_{\intr}}}{A+1}
    - \frac{2}{m A(A+1)} \sum_{\alpha}^{<\epsilon_F} 
      \matrixe{\alpha \beta}{\qOV^2}{\alpha\beta} \;.
}
In addition to the eigenvalues $\epsilon_{\beta}$ of the HF Hamiltonian \eqref{eq:hf_hfhamiltonian}, a global shift depending on the intrinsic kinetic energy of the $A$-body system and a state-dependent correction appear. These relations are be used to define corrected single-particle energies $\epsilon^{\text{corr}}_{\beta}$ which can be compared to single-particle energies extracted from experimental data or other calculations. 

\begin{figure}
\begin{center}
\includegraphics[width=0.95\columnwidth]{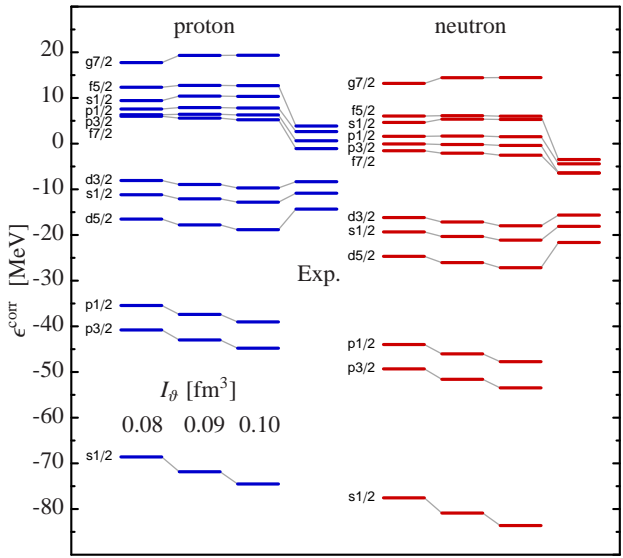}
\end{center}
\caption{(Color online) Single-particle energies for \elem{Ca}{40}, obtained with the correlated AV18 potential using three different values of $I_{\vartheta}$ as indicated by the labels. The experimental data is taken from \cite{IsEr02}.}
\label{fig:hf_spenergies}
\end{figure}

In Fig. \ref{fig:hf_spenergies} the corrected single-particle energies for \elem{Ca}{40} are shown for three different ranges of the triplet-even tensor correlator around the optimal value of $I_{\vartheta}=0.09\,\text{fm}^3$. The overall agreement with the experimental levels is reasonable. The position of the highest occupied single-particle state is reproduced quite well for the optimal correlator range. Nevertheless, the average level spacing seems too large, leading to very deeply bound single-particle states at the lower end of the spectrum. An analogous calculation based on a $V_{\text{low}k}$ effective interaction derived from the AV18 potential has shown the same effect in an even more pronounced way \cite{CoIt03}.

\begin{table}[t]
\caption{Difference of the HF single-particle energies for spin-orbit partner states for protons ($\pi$) and neutrons ($\nu$) in various nuclei, obtained with the correlated AV18 potential for different $I_{\vartheta}$. The experimental values are taken from \cite{IsEr02}.}
\label{tab:hf_spinorbitsplittings}
\begin{ruledtabular}
\begin{tabular}{c c c c c c}
Nucleus & Orbital & \multicolumn{3}{c}{$I_{\vartheta}\;[\text{fm}^3]$} & Exp. \\
        &       & $0.08$ & $0.09$ & $0.10$ &  \\
\hline
\elem{O}{16}   & $\pi\,0p$ &  6.40   &  6.77   &  7.06   & 6.32 \\
               & $\pi\,0d$ &  3.57   &  3.82   &  4.03   & 5.00 \\
               & $\nu\,0p$ &  6.41   &  6.78   &  7.07   & 6.18 \\
               & $\nu\,0d$ &  4.22   &  4.51   &  4.76   & 5.08 \\
\hline
\elem{Ca}{40}  & $\pi\,0d$ &  8.44   &  8.85   &  9.14   & 6.00 \\
               & $\pi\,0f$ &  6.35   &  7.16   &  7.47   & 4.95 \\
               & $\pi\,1p$ &  1.25   &  1.47   &  1.52   & 2.01 \\
               & $\nu\,0d$ &  8.49   &  8.89   &  9.18   & 6.00 \\
               & $\nu\,0f$ &  7.60   &  8.21   &  8.54   & 4.88 \\
               & $\nu\,1p$ &  1.67   &  1.85   &  1.91   & 2.00 \\
\hline
\elem{Sn}{100} & $\pi\,0g$ &  5.35   &  5.68   &  5.90   & 6.82 \\  
               & $\pi\,1p$ &  1.72   &  1.83   &  1.90   & 2.85 \\
               & $\nu\,0g$ &  5.03   &  5.35   &  5.55   & 7.00 \\  
               & $\nu\,1d$ &  2.29   &  2.44   &  2.54   & 1.93 \\
\hline
\elem{Sn}{132} & $\pi\,0g$ &  4.38   &  4.64   &  4.79   & 6.08 \\  
               & $\pi\,1d$ &  2.18   &  2.31   &  2.40   & 1.48 \\
               & $\nu\,0h$ &  6.34   &  6.70   &  6.94   & 6.53 \\
               & $\nu\,1d$ &  2.54   &  2.70   &  2.81   & 1.65 \\         
               & $\nu\,2p$ &  0.67   &  0.70   &  0.72   & 0.81 \\
\hline
\elem{Pb}{208} & $\pi\,1d$ &  2.19   &  2.30   &  2.37   & 1.33 \\
               & $\pi\,0h$ &  4.88   &  5.13   &  5.27   & 5.56 \\ 
               & $\nu\,1f$ &  3.32   &  3.51   &  3.63   & 1.77 \\
               & $\nu\,0i$ &  6.75   &  7.10   &  7.32   & 5.84 \\
               & $\nu\,2p$ &  1.30   &  1.36   &  1.41   & 0.90 \\
\end{tabular}
\end{ruledtabular}
\end{table}

Apart from the overall structure of the single-particle spectrum, the energy splitting between spin-orbit partner states provides some insight into the structure of the correlated interaction. These energy differences should be less sensitive to the gross properties of the single-particle Hamiltonian, but emphasize its spin-orbit structure. In Table \ref{tab:hf_spinorbitsplittings} we report the energy differences for spin-orbit partners in the vicinity of the Fermi energy for different nuclei. The experimental data are taken from \cite{IsEr02}. One should keep in mind that there are sizable differences between different experimental data sets. For example, the experimental estimates for the $0d$ splittings in \elem{Ca}{40} range from 6 to 8 MeV \cite{LoVa00}.
 
The overall agreement of the UCOM-HF results with experimental data is quite good and no systematic deviation is evident. This shows that the spin-orbit structure of the correlated interaction is reasonable. Moreover, the results highlight the role of tensor correlations. In all cases, the spin-orbit splittings increase with increasing range of the tensor correlator.

These results on the single-particle states do not include the effects of long-range correlations and of three-body interactions. This remains an important task for future investigations.

\section{Many-Body Perturbation Theory}
\label{sec:pt}

In order to assess the importance of residual long-range correlations we use the Hartree-Fock solution as the starting point for a perturbative calculation. This will allow us to disentangle the effect of residual correlations, which can be included via perturbation theory, from the effect of missing three-body forces.

\subsection{Formulation}

The results of the Hartree-Fock calculations clearly show the importance of residual long-range correlations for the description of nuclei based on realistic NN-interactions. Since these system-dependent correlations cannot be described by the same unitary transformation employed in order to include the dominant short-range correlations, we have to extend our model-space such that long-range correlations can be described by the many-body states themselves. 

The simplest way to estimate the effect of long-range correlations is many-body perturbation theory (MBPT). Many-body perturbation theory starting from the HF solution is a standard technique in many fields of quantum many-body physics, ranging from quantum chemistry \cite{SzOs96} to nuclear physics \cite{Gold57,FeMa77,StSt01,CoIt03}. It is straightforward to apply but has inherent limitations. It is well known that the convergence of successive orders of perturbation theory is not guaranteed. As soon as there are near-degeneracies in the single-partice spectrum, convergence problems are inevitable \cite{DiSc93}. Nevertheless, low-order MBPT provides a quantitative measure for residual contributions beyond HF due to long-range correlations. Of course, a description of the dominant short-range correlations by means of perturbation theory is not possible---it is crucial that those are treated explicitly by the unitary transformation first.

We will restrict ourselves mainly to second order calculations and use the third order contributions only to estimate higher-order effects. The second order contribution involves antisymmetrized two-body matrix elements of the correlated intrinsic Hamiltonian $\corr{\HO}_{\intr} = \TO_{\intr} + \VO_{\UCOM}$ between two states below the Fermi energy (hole states denoted by $\alpha,\alpha',...$) and two states above the Fermi energy (particle states denoted by $\beta,\beta',...$):
\eq{
  E^{(2)} 
  = \frac{1}{4} 
    \sum_{\alpha,\alpha'}^{<\epsilon_F}  \sum_{\beta,\beta'}^{>\epsilon_F}  
    \frac{|\matrixe{\alpha\alpha'}{\corr{\HO}_{\intr}}{\beta\beta'}|^2}
      {(\epsilon_{\alpha} + \epsilon_{\alpha'} - \epsilon_{\beta} - \epsilon_{\beta'})} \;.
}
Note that the full two-body part of the many-body Hamiltonian enters, which includes the intrinsic kinetic energy in our case.

The third-order contribution can be conveniently decomposed into three parts \cite{StSt01}: One involving two additional particle states,
\eqmulti{
  E^{(3)}_{pp} 
  &= \frac{1}{8} 
    \sum_{\alpha,\alpha'}^{<\epsilon_F}  
    \sum_{\beta\beta'\beta''\beta'''}^{>\epsilon_F} \\
    &\frac{
      \matrixe{\alpha\alpha'}{\corr{\HO}_{\intr}}{\beta\beta'}
      \matrixe{\beta\beta'}{\corr{\HO}_{\intr}}{\beta''\beta'''}
      \matrixe{\beta''\beta'''}{\corr{\HO}_{\intr}}{\alpha\alpha'}
    }{
      (\epsilon_{\alpha} + \epsilon_{\alpha'} - \epsilon_{\beta} - \epsilon_{\beta'})
      (\epsilon_{\alpha} + \epsilon_{\alpha'} - \epsilon_{\beta''} - \epsilon_{\beta'''})
    } \;,
}
one with two additional hole states,
\eqmulti{
  E^{(3)}_{hh} 
  &= \frac{1}{8} 
    \sum_{\alpha\alpha'\alpha''\alpha'''}^{<\epsilon_F}  
    \sum_{\beta\beta'}^{>\epsilon_F}  \\
    &\frac{
      \matrixe{\alpha\alpha'}{\corr{\HO}_{\intr}}{\beta\beta'}
      \matrixe{\beta\beta'}{\corr{\HO}_{\intr}}{\alpha''\alpha'''}
      \matrixe{\alpha''\alpha'''}{\corr{\HO}_{\intr}}{\alpha\alpha'}
    }{
      (\epsilon_{\alpha} + \epsilon_{\alpha'} - \epsilon_{\beta} - \epsilon_{\beta'})
      (\epsilon_{\alpha''} + \epsilon_{\alpha'''} - \epsilon_{\beta} - \epsilon_{\beta'})
    } \;,
}
and a third part with one additional particle and one additional hole state:
\eqmulti{
  E^{(3)}_{ph} 
  &= 
    \sum_{\alpha\alpha'\alpha''}^{<\epsilon_F}   
    \sum_{\beta\beta'\beta''}^{>\epsilon_F}  \\
    &\frac{
      \matrixe{\alpha\alpha'}{\corr{\HO}_{\intr}}{\beta\beta'}
      \matrixe{\alpha''\beta}{\corr{\HO}_{\intr}}{\alpha\beta''}
      \matrixe{\beta'\beta''}{\corr{\HO}_{\intr}}{\alpha''\alpha'}
    }{
      (\epsilon_{\alpha} + \epsilon_{\alpha'} - \epsilon_{\beta} - \epsilon_{\beta'})
      (\epsilon_{\alpha'} + \epsilon_{\alpha''} - \epsilon_{\beta'} - \epsilon_{\beta''})
    } \;.
}
The numerical evaluation of the third order contributions is extremely time-consuming. Moreover, it does not necessarily improve the results nor does it prove convergence \cite{DiSc93}. 

Perturbation theory can also be used to construct the perturbed many-body states, which in turn give access to the other observables. We will not go into detail (see Ref. \cite{StBa73}) but rather present a few results on the effect of second order perturbative corrections on occupation propabilities and charge radii in Sec. \ref{sec:pt_radii}. 

\subsection{Ground-state energies}

For all following calculations we again use the correlated AV18 potential for the triplet-even tensor correlator with the optimal range $I_{\vartheta}=0.09\,\text{fm}^3$, as determined from no-core shell model calculations (cf. Sec. \ref{sec:ucom_optcorr}).

\begin{figure}
\begin{center}
\includegraphics[width=1\columnwidth]{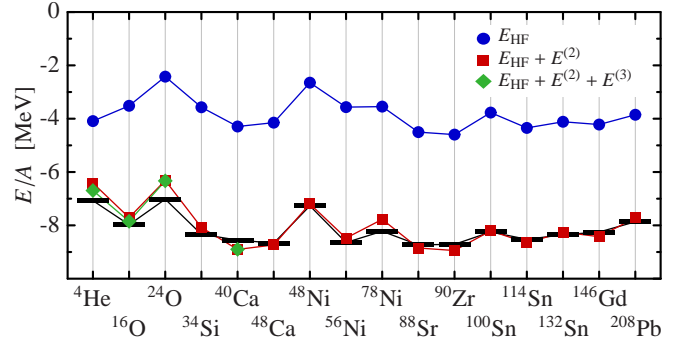}
\end{center}
\caption{(Color online) Ground state energies for selected closed-shell nuclei in HF approximation and with added second and third order MBPT corrections. The correlated AV18 potential with $I_{\vartheta}=0.09\,\text{fm}^3$ was used. The bars indicate the experimental binding energies \cite{AuWa95}.}
\label{fig:pt_stdnucl}
\end{figure}

Figure \ref{fig:pt_stdnucl} compares the ground state energies in HF approximation and second order perturbation theory for selected closed shell nuclei. All calculations were performed using $e_{\max}=12$ major oscillator shells in order to ensure a satisfactory degree of convergence of the perturbative contributions. 
The residual change in binding energy when going from $e_{\max}=12$ to $e_{\max}=13$ is on the level of 3\% for \elem{Ca}{40} and \elem{Zr}{90}.
For light nuclei the third order perturbative contributions are also shown. However, owing to the high computational cost, a reduced basis set with $e_{\max}=8$ was used.

The inclusion of the perturbative contributions to the energy leads to a remarkable result. Throughout the whole mass range, we obtain a good agreement with the experimental binding energies. The binding energy missing in the HF treatment is completely recovered by the second order perturbative contribution $E^{(2)}$. In all cases we considered, the third order contribution $E^{(3)}$ is very small, but tends to improve the agreement with the experimental energies further.

\begin{figure}
\begin{center}
\includegraphics[width=1\columnwidth]{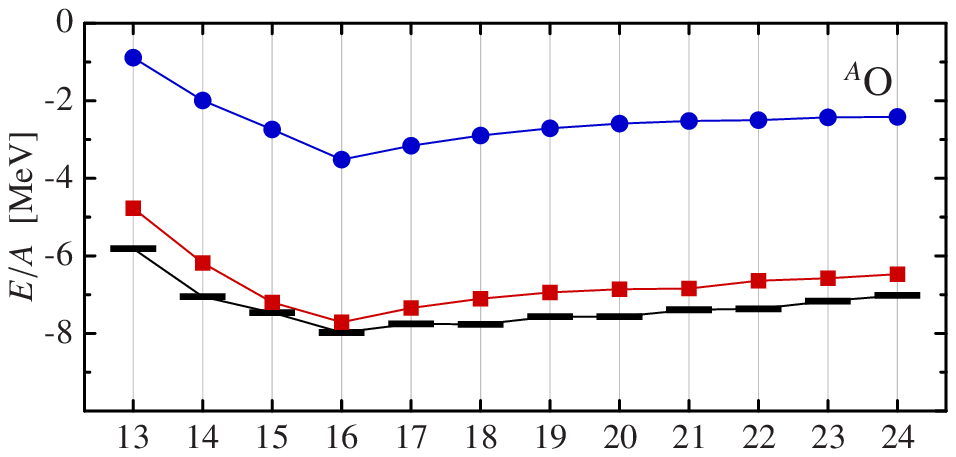}
\includegraphics[width=1\columnwidth]{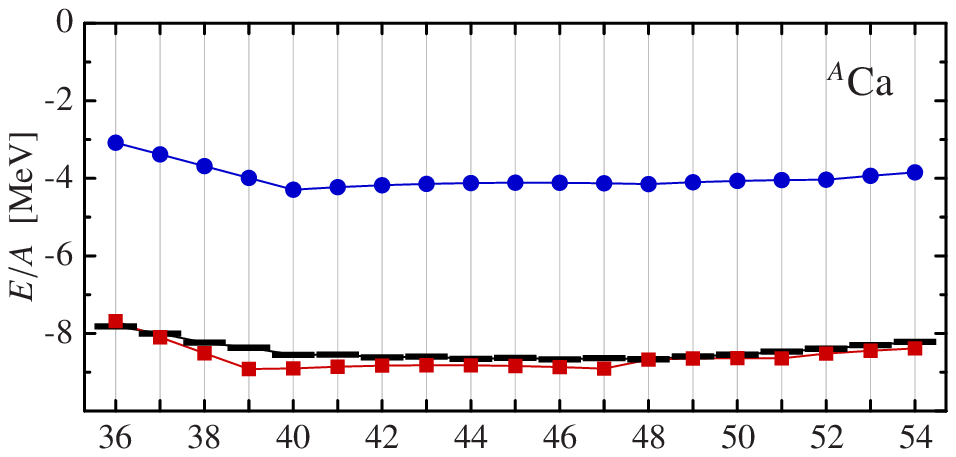}
\includegraphics[width=1\columnwidth]{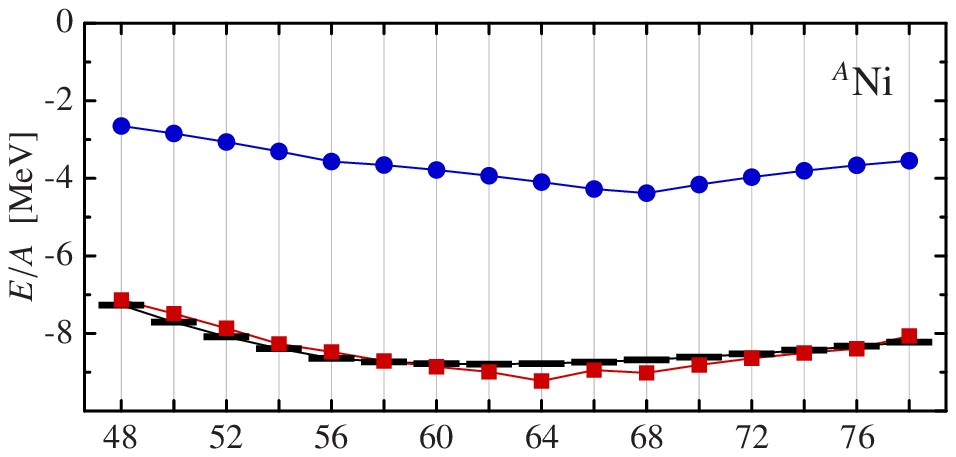}
\includegraphics[width=1\columnwidth]{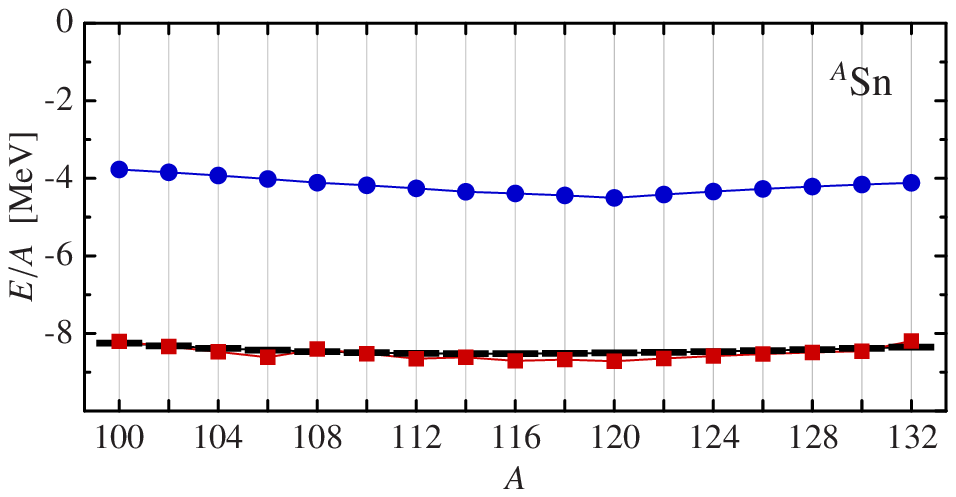}
\end{center}
\caption{(Color online) Ground state energies for the O, Ca, Ni, and Sn isotope chains in HF approximation and with added second order MBPT corrections (see Fig. \ref{fig:pt_stdnucl}). The correlated AV18 potential with $I_{\vartheta}=0.09\,\text{fm}^3$ was used. The bars indicate the experimental binding energies \cite{AuWa95}.}
\label{fig:pt_isochains}
\end{figure}

This observation is also confirmed for open-shell nuclei. We extend the HF and MBPT schemes by allowing for partially filled $nlj$-shells under the constraint of identical single-particle states for each $m$-sublevel (cf. Sec. \ref{sec:hf}). This, of course, does not account for effects like pairing and deformation which will be discussed elsewhere. Nevertheless, it allows us to systematically investigate the isospin-dependence of the correlated interaction. Figure \ref{fig:pt_isochains} shows the HF and the HF+MBPT energies for the O, Ca, Ni, and Sn isotope chains. Again, the agreement of $E_{\text{HF}}+E^{(2)}$ with the experimental ground state energies is remarkable, even for extreme neutron numbers. This shows that the isospin-character of the correlated interaction is realistic and assures predictive power also far off the valley of stability.   

These results entail two important conclusions: (\emph{i}) The residual long-range correlations behave perturbatively and can be described well within MBPT. The essential step of taming the realistic NN-interaction with regard to the strong non-perturbative short-range central and tensor correlations was accomplished within the UCOM framework by the unitary transformation. This is encouraging, also in view of more refined methods of extending the model space beyond the HF Slater determinant, e.g., shell-model or configuration interaction (CI) and coupled cluster (CC) calculations \cite{DeHj04}. In a forthcoming publication, we will compare the MBPT results with explicit CI calculations based on $\VO_{\UCOM}$. 

(\emph{ii}) Considering ground state energies only, it seems that the cancellation between genuine three-body forces and the omitted three-body contribution of the cluster expansion of the correlated Hamiltonian works nicely throughout the nuclear chart. If residual long-range correlations are included by means of MBPT, then the experimental binding energies are reproduced without systematic deviations, leaving no room for a net contribution of the three-body force to the energy. However, this might be different for other observables.

\subsection{Occupation probabilities \& charge radii}
\label{sec:pt_radii}

Based on the perturbed many-body state we can study the impact of the residual correlations on other quantities of interest. Here we will restrict ourselves to two aspects: First, the change of the occupation probabilities of the single-particle orbitals as a probe for the structure of the perturbed state. Second, the charge rms-radius as a global indicator for the change of the density distribution.    

We adopt the formulation of the perturbative corrections to the one-body density matrix given in Ref. \cite{StBa73}. The matrix elements of the perturbed density matrix in the HF single-particle basis are constructed from the perturbed many-body states including all contributions up to second order in the perturbation. The diagonal matrix elements directly provide the mean occupation numbers $\bar{n}_{\alpha}$ of the HF single-particle states.

\begin{figure}
\begin{center}
\includegraphics[width=1\columnwidth]{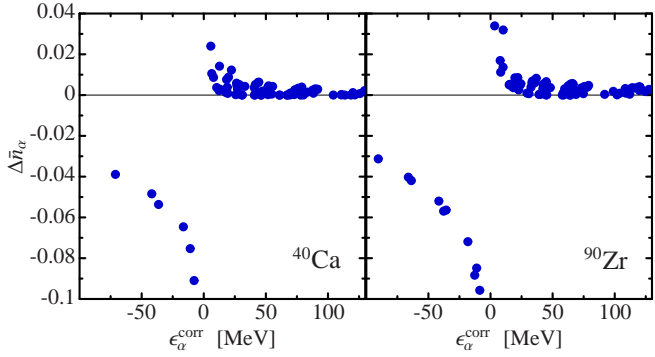}
\end{center}
\caption{(Color online) Change of the mean occupation numbers of the HF single-particle states due to second order perturbative corrections. Shown are the results for proton orbitals in \elem{Ca}{40} and \elem{Zr}{90} as functions of the corrected single-particle energy. The correlated AV18 potential with $I_{\vartheta}=0.09\,\text{fm}^3$ was used.}
\label{fig:pt_occu}
\end{figure}

The changes in the occupation numbers of the proton states in \elem{Ca}{40} and \elem{Zr}{90} are depicted in Fig. \ref{fig:pt_occu}. The occupation of levels below the Fermi energy is depleted and levels above the Fermi energy are populated. Just below the Fermi energy the depletion can reach up to $10\%$. The total depletion of the proton states below the Fermi energy is between $6$ and $7\%$ for both nuclei. The population of states right above the Fermi energy reaches approximately $4\%$. With increasing single-particle energy the population of the particle states deceases rapidly and becomes rather small for the largest energies contained in the single-particle space. 

One should keep in mind that these results only reflect the impact of the long-range correlations treated by perturbation theory. The dominant short-range correlations, which are described by the unitary correlators, do not show up in these occupation numbers. In order to reveal their impact as well, one has to formulate a correlated occupation number operator, e.g., with respect to momentum eigenstates.  

\begin{figure}
\begin{center}
\includegraphics[width=1\columnwidth]{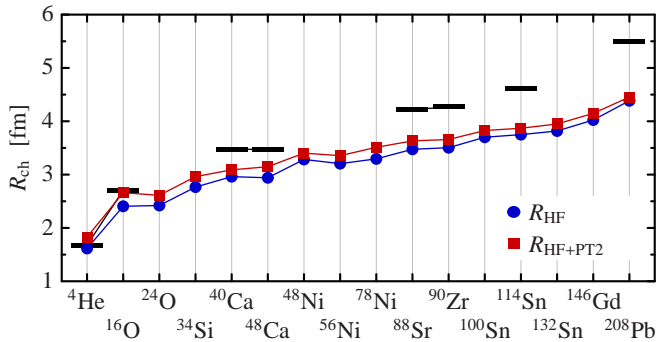}
\end{center}
\caption{(Color online) Charge radii for selected closed-shell nuclei in HF approximation and with added second MBPT corrections. The correlated AV18 potential with $I_{\vartheta}=0.09\,\text{fm}^3$ was used. The bars indicate experimental charge radii \cite{VrJa87}.}
\label{fig:pt_radii}
\end{figure}

By contracting the one-body density matrix with the wave functions of the HF single-particle states we determine the perturbed proton and neutron density distributions. Charge distributions and charge radii are obtained including the proton and neutron form factors as well as a center of mass correction. The perturbed charge radii of closed shell nuclei are summarized in Fig. \ref{fig:pt_radii}. The perturbative corrections increase the charge radii by typically $0.1\,\text{fm}$ to $0.2\,\text{fm}$. The increase is the result of individual contributions of different signs which could also cause a decrease of the radius \cite{StBa73}. This result is consistent with the general expectation that the admixture of higher-lying states increases the radii. However, the observed increase is not sufficient to obtain agreement with the available experimental data for heavier nuclei. With growing mass number, the deviation from the experimental radii increases. 

Assuming the validity of the perturbative estimate, this implies that the deviation of the HF charge radii from the experimental ones cannot be fully explained by long-range correlations. Hence, it can be interpreted as an indication for the necessity of a net effective three-body force, i.e. a combination of the genuine three-body force and the three-body contributions of the cluster expansion.

\section{Conclusions}

We have employed the unitary correlation operator method for describing the dominant short-range correlations induced by realistic NN-potentials in a simple Hartree-Fock framework. Based on the Argonne V18 potential with optimal correlation functions determined in the two-body system and a range constraint for the tensor correlation functions fixed in three- and four-body systems, we have performed HF and MBPT calculations for spherical nuclei throughout the nuclear chart.  

We obtain bound nuclei using the correlated AV18 potential already at the HF level. This proves that the dominant short-range central and tensor correlations are successfully described by the unitary correlation operators. Without the proper inclusion of both types of correlations it is not possible to obtain self-bound solutions in a HF framework using the AV18 potential. However, the HF binding energies remain significantly smaller than the experimental binding energies. The same holds true for charge radii. On the other hand, the single-particle energy differences between spin-orbit partner states show a satisfactory agreement with experimental estimates.

The missing binding energy is connected to residual long-range correlations, which are not described by the unitary correlation operators. They have to be covered by the model space and the Slater determinant of the HF approximation is clearly not able to do so. Many-body perturbation theory as the simplest possible step beyond the HF ground state already recovers the missing binding energy. The agreement between second order ground state energies and experimental data is remarkably good throughout the whole mass range from \elem{He}{4} to \elem{Pb}{208}, even far off the valley of stability. Unlike the short-range central and tensor correlations, the residual long-range correlations are perturbative. This opens interesting perspectives for the application of more refined many-body techniques, like configuration interaction and coupled-cluster schemes, to benchmark the perturbative results and obtain a more detailed insight into the structure of those correlations.
 
None of the calculations presented here does include three-body forces. Therefore, it is surprising that a good agreement with the experimental binding energies was observed for all nuclei considered. This is due to a net cancellation of the energy contributions of the genuine three-body force (attractive) and the three-body order of the cluster expansion (repulsive). This was already observed in no-core shell model calculations for light systems \cite{RoHe05}, but seems to hold across the whole nuclear chart. Obviously, this cancellation effect does not necessarily work for other observables as well. The charge radii, which still show a sizable deviation from experiment after including long-range correlations, point into that direction. The construction and inclusion of effective three-body forces will therefore be one of the major lines of research for the future.

\section*{Acknowledgments}

This work is supported by the Deutsche Forschungsgemeinschaft through contract SFB 634. We thank the Institute for Nuclear Theory at the University of Washington for its hospitality and the Department of Energy for partial support during the completion of this work.


\begin{thebibliography}{43}
\expandafter\ifx\csname natexlab\endcsname\relax\def\natexlab#1{#1}\fi
\expandafter\ifx\csname bibnamefont\endcsname\relax
  \def\bibnamefont#1{#1}\fi
\expandafter\ifx\csname bibfnamefont\endcsname\relax
  \def\bibfnamefont#1{#1}\fi
\expandafter\ifx\csname citenamefont\endcsname\relax
  \def\citenamefont#1{#1}\fi
\expandafter\ifx\csname url\endcsname\relax
  \def\url#1{\texttt{#1}}\fi
\expandafter\ifx\csname urlprefix\endcsname\relax\def\urlprefix{URL }\fi
\providecommand{\bibinfo}[2]{#2}
\providecommand{\eprint}[2][]{\url{#2}}

\bibitem[{\citenamefont{Finelli et~al.}(2004)\citenamefont{Finelli, Kaiser,
  Vretenar, and Weise}}]{FiKa04}
\bibinfo{author}{\bibfnamefont{P.}~\bibnamefont{Finelli}},
  \bibinfo{author}{\bibfnamefont{N.}~\bibnamefont{Kaiser}},
  \bibinfo{author}{\bibfnamefont{D.}~\bibnamefont{Vretenar}}, \bibnamefont{and}
  \bibinfo{author}{\bibfnamefont{W.}~\bibnamefont{Weise}},
  \bibinfo{journal}{Nucl. Phys.} \textbf{\bibinfo{volume}{A735}},
  \bibinfo{pages}{449} (\bibinfo{year}{2004}).

\bibitem[{\citenamefont{Pieper and Wiringa}(2001)}]{PiWi01}
\bibinfo{author}{\bibfnamefont{S.~C.} \bibnamefont{Pieper}} \bibnamefont{and}
  \bibinfo{author}{\bibfnamefont{R.~B.} \bibnamefont{Wiringa}},
  \bibinfo{journal}{Ann. Rev. Nucl. Part. Sci.} \textbf{\bibinfo{volume}{51}},
  \bibinfo{pages}{53} (\bibinfo{year}{2001}).

\bibitem[{\citenamefont{Pieper et~al.}(2004)\citenamefont{Pieper, Wiringa, and
  Carlson}}]{PiWi04}
\bibinfo{author}{\bibfnamefont{S.~C.} \bibnamefont{Pieper}},
  \bibinfo{author}{\bibfnamefont{R.~B.} \bibnamefont{Wiringa}},
  \bibnamefont{and} \bibinfo{author}{\bibfnamefont{J.}~\bibnamefont{Carlson}},
  \bibinfo{journal}{Phys. Rev. C} \textbf{\bibinfo{volume}{70}},
  \bibinfo{pages}{054325} (\bibinfo{year}{2004}).

\bibitem[{\citenamefont{Pieper et~al.}(2002)\citenamefont{Pieper, Varga, and
  Wiringa}}]{PiVa02}
\bibinfo{author}{\bibfnamefont{S.~C.} \bibnamefont{Pieper}},
  \bibinfo{author}{\bibfnamefont{K.}~\bibnamefont{Varga}}, \bibnamefont{and}
  \bibinfo{author}{\bibfnamefont{R.~B.} \bibnamefont{Wiringa}},
  \bibinfo{journal}{Phys. Rev. C} \textbf{\bibinfo{volume}{66}},
  \bibinfo{pages}{044310} (\bibinfo{year}{2002}).

\bibitem[{\citenamefont{Caurier et~al.}(2002)\citenamefont{Caurier, Navr\'atil,
  Ormand, and Vary}}]{CaNa02}
\bibinfo{author}{\bibfnamefont{E.}~\bibnamefont{Caurier}},
  \bibinfo{author}{\bibfnamefont{P.}~\bibnamefont{Navr\'atil}},
  \bibinfo{author}{\bibfnamefont{W.~E.} \bibnamefont{Ormand}},
  \bibnamefont{and} \bibinfo{author}{\bibfnamefont{J.~P.} \bibnamefont{Vary}},
  \bibinfo{journal}{Phys. Rev. C} \textbf{\bibinfo{volume}{66}},
  \bibinfo{pages}{024314} (\bibinfo{year}{2002}).

\bibitem[{\citenamefont{Navr\'atil and Ormand}(2002)}]{NaOr02}
\bibinfo{author}{\bibfnamefont{P.}~\bibnamefont{Navr\'atil}} \bibnamefont{and}
  \bibinfo{author}{\bibfnamefont{W.~E.} \bibnamefont{Ormand}},
  \bibinfo{journal}{Phys. Rev. Lett.} \textbf{\bibinfo{volume}{88}},
  \bibinfo{pages}{152502} (\bibinfo{year}{2002}).

\bibitem[{\citenamefont{Navr\'atil et~al.}(2000)\citenamefont{Navr\'atil, Vary,
  and Barrett}}]{NaVa00}
\bibinfo{author}{\bibfnamefont{P.}~\bibnamefont{Navr\'atil}},
  \bibinfo{author}{\bibfnamefont{J.~P.} \bibnamefont{Vary}}, \bibnamefont{and}
  \bibinfo{author}{\bibfnamefont{B.~R.} \bibnamefont{Barrett}},
  \bibinfo{journal}{Phys. Rev. Lett.} \textbf{\bibinfo{volume}{84}},
  \bibinfo{pages}{5728} (\bibinfo{year}{2000}).

\bibitem[{\citenamefont{Wiringa et~al.}(1995)\citenamefont{Wiringa, Stoks, and
  Schiavilla}}]{WiSt95}
\bibinfo{author}{\bibfnamefont{R.~B.} \bibnamefont{Wiringa}},
  \bibinfo{author}{\bibfnamefont{V.~G.~J.} \bibnamefont{Stoks}},
  \bibnamefont{and}
  \bibinfo{author}{\bibfnamefont{R.}~\bibnamefont{Schiavilla}},
  \bibinfo{journal}{Phys. Rev. C} \textbf{\bibinfo{volume}{51}},
  \bibinfo{pages}{38} (\bibinfo{year}{1995}).

\bibitem[{\citenamefont{Machleidt}(2001)}]{Mach01}
\bibinfo{author}{\bibfnamefont{R.}~\bibnamefont{Machleidt}},
  \bibinfo{journal}{Phys. Rev. C} \textbf{\bibinfo{volume}{63}},
  \bibinfo{pages}{024001} (\bibinfo{year}{2001}).

\bibitem[{\citenamefont{Feldmeier et~al.}(1998)\citenamefont{Feldmeier, Neff,
  Roth, and Schnack}}]{FeNe98}
\bibinfo{author}{\bibfnamefont{H.}~\bibnamefont{Feldmeier}},
  \bibinfo{author}{\bibfnamefont{T.}~\bibnamefont{Neff}},
  \bibinfo{author}{\bibfnamefont{R.}~\bibnamefont{Roth}}, \bibnamefont{and}
  \bibinfo{author}{\bibfnamefont{J.}~\bibnamefont{Schnack}},
  \bibinfo{journal}{Nucl. Phys.} \textbf{\bibinfo{volume}{A632}},
  \bibinfo{pages}{61} (\bibinfo{year}{1998}).

\bibitem[{\citenamefont{Neff and Feldmeier}(2003)}]{NeFe03}
\bibinfo{author}{\bibfnamefont{T.}~\bibnamefont{Neff}} \bibnamefont{and}
  \bibinfo{author}{\bibfnamefont{H.}~\bibnamefont{Feldmeier}},
  \bibinfo{journal}{Nucl. Phys.} \textbf{\bibinfo{volume}{A713}},
  \bibinfo{pages}{311} (\bibinfo{year}{2003}).

\bibitem[{\citenamefont{Roth et~al.}(2004)\citenamefont{Roth, Neff, Hergert,
  and Feldmeier}}]{RoNe04}
\bibinfo{author}{\bibfnamefont{R.}~\bibnamefont{Roth}},
  \bibinfo{author}{\bibfnamefont{T.}~\bibnamefont{Neff}},
  \bibinfo{author}{\bibfnamefont{H.}~\bibnamefont{Hergert}}, \bibnamefont{and}
  \bibinfo{author}{\bibfnamefont{H.}~\bibnamefont{Feldmeier}},
  \bibinfo{journal}{Nucl. Phys.} \textbf{\bibinfo{volume}{A745}},
  \bibinfo{pages}{3} (\bibinfo{year}{2004}).

\bibitem[{\citenamefont{Roth et~al.}(2005)\citenamefont{Roth, Hergert,
  Papakonstantinou, Neff, and Feldmeier}}]{RoHe05}
\bibinfo{author}{\bibfnamefont{R.}~\bibnamefont{Roth}},
  \bibinfo{author}{\bibfnamefont{H.}~\bibnamefont{Hergert}},
  \bibinfo{author}{\bibfnamefont{P.}~\bibnamefont{Papakonstantinou}},
  \bibinfo{author}{\bibfnamefont{T.}~\bibnamefont{Neff}}, \bibnamefont{and}
  \bibinfo{author}{\bibfnamefont{H.}~\bibnamefont{Feldmeier}},
  \bibinfo{journal}{Phys. Rev. C} \textbf{\bibinfo{volume}{72}},
  \bibinfo{pages}{034002} (\bibinfo{year}{2005}).

\bibitem[{\citenamefont{Day}(1967)}]{Day67}
\bibinfo{author}{\bibfnamefont{B.}~\bibnamefont{Day}}, \bibinfo{journal}{Rev.
  Mod. Phys.} \textbf{\bibinfo{volume}{39}}, \bibinfo{pages}{719}
  (\bibinfo{year}{1967}).

\bibitem[{\citenamefont{Bogner et~al.}(2003{\natexlab{a}})\citenamefont{Bogner,
  Kuo, and Schwenk}}]{BoKu03}
\bibinfo{author}{\bibfnamefont{S.~K.} \bibnamefont{Bogner}},
  \bibinfo{author}{\bibfnamefont{T.~T.~S.} \bibnamefont{Kuo}},
  \bibnamefont{and} \bibinfo{author}{\bibfnamefont{A.}~\bibnamefont{Schwenk}},
  \bibinfo{journal}{Phys. Rep.} \textbf{\bibinfo{volume}{386}},
  \bibinfo{pages}{1} (\bibinfo{year}{2003}{\natexlab{a}}).

\bibitem[{\citenamefont{Bogner et~al.}(2003{\natexlab{b}})\citenamefont{Bogner,
  Kuo, Schwenk, Entem, and Machleidt}}]{BoKu03b}
\bibinfo{author}{\bibfnamefont{S.~K.} \bibnamefont{Bogner}},
  \bibinfo{author}{\bibfnamefont{T.~T.~S.} \bibnamefont{Kuo}},
  \bibinfo{author}{\bibfnamefont{A.}~\bibnamefont{Schwenk}},
  \bibinfo{author}{\bibfnamefont{D.~R.} \bibnamefont{Entem}}, \bibnamefont{and}
  \bibinfo{author}{\bibfnamefont{R.}~\bibnamefont{Machleidt}},
  \bibinfo{journal}{Phys. Lett.} \textbf{\bibinfo{volume}{B576}},
  \bibinfo{pages}{265} (\bibinfo{year}{2003}{\natexlab{b}}).

\bibitem[{\citenamefont{Suzuki and Lee}(1980)}]{SuLe80}
\bibinfo{author}{\bibfnamefont{K.}~\bibnamefont{Suzuki}} \bibnamefont{and}
  \bibinfo{author}{\bibfnamefont{S.~Y.} \bibnamefont{Lee}},
  \bibinfo{journal}{Prog. Theo. Phys.} \textbf{\bibinfo{volume}{64}},
  \bibinfo{pages}{2091} (\bibinfo{year}{1980}).

\bibitem[{\citenamefont{Fujii et~al.}(2004)\citenamefont{Fujii, Okamoto, and
  Suzuki}}]{FuOk04}
\bibinfo{author}{\bibfnamefont{S.}~\bibnamefont{Fujii}},
  \bibinfo{author}{\bibfnamefont{R.}~\bibnamefont{Okamoto}}, \bibnamefont{and}
  \bibinfo{author}{\bibfnamefont{K.}~\bibnamefont{Suzuki}},
  \bibinfo{journal}{Phys. Rev. C} \textbf{\bibinfo{volume}{69}},
  \bibinfo{pages}{034328} (\bibinfo{year}{2004}).

\bibitem[{\citenamefont{Suzuki et~al.}(1994)\citenamefont{Suzuki, Okamoto, and
  Kumagai}}]{SuOk94}
\bibinfo{author}{\bibfnamefont{K.}~\bibnamefont{Suzuki}},
  \bibinfo{author}{\bibfnamefont{R.}~\bibnamefont{Okamoto}}, \bibnamefont{and}
  \bibinfo{author}{\bibfnamefont{H.}~\bibnamefont{Kumagai}},
  \bibinfo{journal}{Phys. Rep.} \textbf{\bibinfo{volume}{242}},
  \bibinfo{pages}{181} (\bibinfo{year}{1994}).

\bibitem[{\citenamefont{Shakin et~al.}(1967)\citenamefont{Shakin, Waghmare, and
  Hull}}]{ShWa67}
\bibinfo{author}{\bibfnamefont{C.~M.} \bibnamefont{Shakin}},
  \bibinfo{author}{\bibfnamefont{Y.~R.} \bibnamefont{Waghmare}},
  \bibnamefont{and} \bibinfo{author}{\bibfnamefont{M.~H.} \bibnamefont{Hull}},
  \bibinfo{journal}{Phys. Rev.} \textbf{\bibinfo{volume}{161}},
  \bibinfo{pages}{1006} (\bibinfo{year}{1967}).

\bibitem[{\citenamefont{Feldmeier}(1990)}]{Feld90}
\bibinfo{author}{\bibfnamefont{H.}~\bibnamefont{Feldmeier}},
  \bibinfo{journal}{Nucl. Phys.} \textbf{\bibinfo{volume}{A515}},
  \bibinfo{pages}{147} (\bibinfo{year}{1990}).

\bibitem[{\citenamefont{Feldmeier and Schnack}(2000)}]{FeSc00}
\bibinfo{author}{\bibfnamefont{H.}~\bibnamefont{Feldmeier}} \bibnamefont{and}
  \bibinfo{author}{\bibfnamefont{J.}~\bibnamefont{Schnack}},
  \bibinfo{journal}{Rev. Mod. Phys.} \textbf{\bibinfo{volume}{72}},
  \bibinfo{pages}{655} (\bibinfo{year}{2000}).

\bibitem[{\citenamefont{Talmi}(1952)}]{Talm52}
\bibinfo{author}{\bibfnamefont{I.}~\bibnamefont{Talmi}},
  \bibinfo{journal}{Helv. Phys. Acta} \textbf{\bibinfo{volume}{25}},
  \bibinfo{pages}{185} (\bibinfo{year}{1952}).

\bibitem[{\citenamefont{Moshinsky}(1959)}]{Mosh59}
\bibinfo{author}{\bibfnamefont{M.}~\bibnamefont{Moshinsky}},
  \bibinfo{journal}{Nucl. Phys.} \textbf{\bibinfo{volume}{13}},
  \bibinfo{pages}{104} (\bibinfo{year}{1959}).

\bibitem[{\citenamefont{Kamuntavicius et~al.}(2001)\citenamefont{Kamuntavicius,
  Kalinauskas, Barrett, Mickevicius, and Germanas}}]{KaKa01}
\bibinfo{author}{\bibfnamefont{G.~P.} \bibnamefont{Kamuntavicius}},
  \bibinfo{author}{\bibfnamefont{R.~K.} \bibnamefont{Kalinauskas}},
  \bibinfo{author}{\bibfnamefont{B.~R.} \bibnamefont{Barrett}},
  \bibinfo{author}{\bibfnamefont{S.}~\bibnamefont{Mickevicius}},
  \bibnamefont{and} \bibinfo{author}{\bibfnamefont{D.}~\bibnamefont{Germanas}},
  \bibinfo{journal}{Nucl. Phys.} \textbf{\bibinfo{volume}{A695}},
  \bibinfo{pages}{191} (\bibinfo{year}{2001}).

\bibitem[{\citenamefont{Buck and Merchant}(1996)}]{BuMe96}
\bibinfo{author}{\bibfnamefont{B.}~\bibnamefont{Buck}} \bibnamefont{and}
  \bibinfo{author}{\bibfnamefont{A.~C.} \bibnamefont{Merchant}},
  \bibinfo{journal}{Nucl. Phys.} \textbf{\bibinfo{volume}{A600}},
  \bibinfo{pages}{387} (\bibinfo{year}{1996}).

\bibitem[{\citenamefont{Ring and Schuck}(1980)}]{RiSc80}
\bibinfo{author}{\bibfnamefont{P.}~\bibnamefont{Ring}} \bibnamefont{and}
  \bibinfo{author}{\bibfnamefont{P.}~\bibnamefont{Schuck}},
  \emph{\bibinfo{title}{The Nuclear Many-Body Problem}}
  (\bibinfo{publisher}{Springer Verlag, New York}, \bibinfo{year}{1980}).

\bibitem[{\citenamefont{Schmid}(2002)}]{Schm02}
\bibinfo{author}{\bibfnamefont{K.~W.} \bibnamefont{Schmid}},
  \bibinfo{journal}{Eur. Phys. J. A} \textbf{\bibinfo{volume}{14}},
  \bibinfo{pages}{413} (\bibinfo{year}{2002}).

\bibitem[{\citenamefont{Rodr\'{\i}guez-Guzm\'an and Schmid}(2004)}]{RoSc04}
\bibinfo{author}{\bibfnamefont{R.~R.} \bibnamefont{Rodr\'{\i}guez-Guzm\'an}}
  \bibnamefont{and} \bibinfo{author}{\bibfnamefont{K.~W.}
  \bibnamefont{Schmid}}, \bibinfo{journal}{Eur. Phys. J. A}
  \textbf{\bibinfo{volume}{19}}, \bibinfo{pages}{45} (\bibinfo{year}{2004}).

\bibitem[{\citenamefont{Audi and Wapstra}(1995)}]{AuWa95}
\bibinfo{author}{\bibfnamefont{G.}~\bibnamefont{Audi}} \bibnamefont{and}
  \bibinfo{author}{\bibfnamefont{A.}~\bibnamefont{Wapstra}},
  \bibinfo{journal}{Nucl. Phys.} \textbf{\bibinfo{volume}{A595}},
  \bibinfo{pages}{409} (\bibinfo{year}{1995}).

\bibitem[{\citenamefont{Vries et~al.}(1987)\citenamefont{Vries, Jager, and
  Vries}}]{VrJa87}
\bibinfo{author}{\bibfnamefont{H.~d.} \bibnamefont{Vries}},
  \bibinfo{author}{\bibfnamefont{C.~W.~d.} \bibnamefont{Jager}},
  \bibnamefont{and} \bibinfo{author}{\bibfnamefont{C.~d.} \bibnamefont{Vries}},
  \bibinfo{journal}{At. Data Nucl. Data Tables} \textbf{\bibinfo{volume}{36}},
  \bibinfo{pages}{495} (\bibinfo{year}{1987}).

\bibitem[{\citenamefont{Khadkikar and Kamble}(1974)}]{KhKa74}
\bibinfo{author}{\bibfnamefont{S.~B.} \bibnamefont{Khadkikar}}
  \bibnamefont{and} \bibinfo{author}{\bibfnamefont{V.~B.}
  \bibnamefont{Kamble}}, \bibinfo{journal}{Nucl. Phys.}
  \textbf{\bibinfo{volume}{A225}}, \bibinfo{pages}{352} (\bibinfo{year}{1974}).

\bibitem[{\citenamefont{Jaqua et~al.}(1992)\citenamefont{Jaqua, Hasan, Vary,
  and Barrett}}]{JaHa92}
\bibinfo{author}{\bibfnamefont{L.}~\bibnamefont{Jaqua}},
  \bibinfo{author}{\bibfnamefont{M.~A.} \bibnamefont{Hasan}},
  \bibinfo{author}{\bibfnamefont{J.~P.} \bibnamefont{Vary}}, \bibnamefont{and}
  \bibinfo{author}{\bibfnamefont{B.~R.} \bibnamefont{Barrett}},
  \bibinfo{journal}{Phys. Rev. C} \textbf{\bibinfo{volume}{46}},
  \bibinfo{pages}{2333} (\bibinfo{year}{1992}).

\bibitem[{\citenamefont{Isakov et~al.}(2002)\citenamefont{Isakov, Erokhina,
  Mach, Sanchez-Vega, and Fogelberg}}]{IsEr02}
\bibinfo{author}{\bibfnamefont{V.~I.} \bibnamefont{Isakov}},
  \bibinfo{author}{\bibfnamefont{K.~I.} \bibnamefont{Erokhina}},
  \bibinfo{author}{\bibfnamefont{H.}~\bibnamefont{Mach}},
  \bibinfo{author}{\bibfnamefont{M.}~\bibnamefont{Sanchez-Vega}},
  \bibnamefont{and}
  \bibinfo{author}{\bibfnamefont{B.}~\bibnamefont{Fogelberg}},
  \bibinfo{journal}{Eur. Phys. J. A} \textbf{\bibinfo{volume}{14}},
  \bibinfo{pages}{29} (\bibinfo{year}{2002}).

\bibitem[{\citenamefont{Coraggio et~al.}(2003)\citenamefont{Coraggio, Itaco,
  Covello, Gargano, and Kuo}}]{CoIt03}
\bibinfo{author}{\bibfnamefont{L.}~\bibnamefont{Coraggio}},
  \bibinfo{author}{\bibfnamefont{N.}~\bibnamefont{Itaco}},
  \bibinfo{author}{\bibfnamefont{A.}~\bibnamefont{Covello}},
  \bibinfo{author}{\bibfnamefont{A.}~\bibnamefont{Gargano}}, \bibnamefont{and}
  \bibinfo{author}{\bibfnamefont{T.~T.~S.} \bibnamefont{Kuo}},
  \bibinfo{journal}{Phys. Rev. C} \textbf{\bibinfo{volume}{68}},
  \bibinfo{pages}{034320} (\bibinfo{year}{2003}).

\bibitem[{\citenamefont{L\'opez-Quelle
  et~al.}(2000)\citenamefont{L\'opez-Quelle, Van~Giai, Marcos, and
  Savushkin}}]{LoVa00}
\bibinfo{author}{\bibfnamefont{M.}~\bibnamefont{L\'opez-Quelle}},
  \bibinfo{author}{\bibfnamefont{N.}~\bibnamefont{Van~Giai}},
  \bibinfo{author}{\bibfnamefont{S.}~\bibnamefont{Marcos}}, \bibnamefont{and}
  \bibinfo{author}{\bibfnamefont{L.~N.} \bibnamefont{Savushkin}},
  \bibinfo{journal}{Phys. Rev. C} \textbf{\bibinfo{volume}{61}},
  \bibinfo{pages}{064321} (\bibinfo{year}{2000}).

\bibitem[{\citenamefont{Szabo and Ostlund}(1996)}]{SzOs96}
\bibinfo{author}{\bibfnamefont{A.}~\bibnamefont{Szabo}} \bibnamefont{and}
  \bibinfo{author}{\bibfnamefont{N.~S.} \bibnamefont{Ostlund}},
  \emph{\bibinfo{title}{Modern Quantum Chemistry}} (\bibinfo{publisher}{Dover
  Publications}, \bibinfo{address}{Mineaola, New York}, \bibinfo{year}{1996}).

\bibitem[{\citenamefont{Goldstone}(1957)}]{Gold57}
\bibinfo{author}{\bibfnamefont{J.}~\bibnamefont{Goldstone}},
  \bibinfo{journal}{Proc. R. Soc. London, Ser. A}
  \textbf{\bibinfo{volume}{239}}, \bibinfo{pages}{267} (\bibinfo{year}{1957}).

\bibitem[{\citenamefont{Stevenson et~al.}(2001)\citenamefont{Stevenson,
  Strayer, and Stone}}]{StSt01}
\bibinfo{author}{\bibfnamefont{P.}~\bibnamefont{Stevenson}},
  \bibinfo{author}{\bibfnamefont{M.~R.} \bibnamefont{Strayer}},
  \bibnamefont{and}
  \bibinfo{author}{\bibfnamefont{J.~R.}~\bibnamefont{Stone}},
  \bibinfo{journal}{Phys. Rev. C} \textbf{\bibinfo{volume}{63}},
  \bibinfo{pages}{054309} (\bibinfo{year}{2001}).

\bibitem[{\citenamefont{Feldmeier and Manakos}(1977)}]{FeMa77}
\bibinfo{author}{\bibfnamefont{H.}~\bibnamefont{Feldmeier}} \bibnamefont{and}
  \bibinfo{author}{\bibfnamefont{P.}~\bibnamefont{Manakos}},
  \bibinfo{journal}{Z. Physik} \textbf{\bibinfo{volume}{281}},
  \bibinfo{pages}{379} (\bibinfo{year}{1977}).

\bibitem[{\citenamefont{Dietz et~al.}(1993)\citenamefont{Dietz, Schmidt,
  Warken, and He\ss{}}}]{DiSc93}
\bibinfo{author}{\bibfnamefont{K.}~\bibnamefont{Dietz}},
  \bibinfo{author}{\bibfnamefont{C.}~\bibnamefont{Schmidt}},
  \bibinfo{author}{\bibfnamefont{M.}~\bibnamefont{Warken}}, \bibnamefont{and}
  \bibinfo{author}{\bibfnamefont{B.~A.} \bibnamefont{He\ss{}}},
  \bibinfo{journal}{J. Phys. B: At. Mol. Opt. Phys}
  \textbf{\bibinfo{volume}{26}}, \bibinfo{pages}{1885} (\bibinfo{year}{1993}).

\bibitem[{\citenamefont{Strayer et~al.}(1973)\citenamefont{Strayer, Bassichis,
  and Kerman}}]{StBa73}
\bibinfo{author}{\bibfnamefont{M.~R.} \bibnamefont{Strayer}},
  \bibinfo{author}{\bibfnamefont{W.~H.} \bibnamefont{Bassichis}},
  \bibnamefont{and} \bibinfo{author}{\bibfnamefont{A.~K.}
  \bibnamefont{Kerman}}, \bibinfo{journal}{Phys. Rev. C}
  \textbf{\bibinfo{volume}{8}}, \bibinfo{pages}{1269} (\bibinfo{year}{1973}).

\bibitem[{\citenamefont{Dean and Hjorth-Jensen}(2004)}]{DeHj04}
\bibinfo{author}{\bibfnamefont{D.~J.} \bibnamefont{Dean}} \bibnamefont{and}
  \bibinfo{author}{\bibfnamefont{M.}~\bibnamefont{Hjorth-Jensen}},
  \bibinfo{journal}{Phys. Rev. C} \textbf{\bibinfo{volume}{69}},
  \bibinfo{pages}{054320} (\bibinfo{year}{2004}).

\end{thebibliography}

\end{document}